\documentstyle[12pt,epsfig]{article}

\def\apm#1{\hbox{$\pm #1$}}
\def\epm#1#2{\hbox{${\lower1pt\hbox{$\scriptstyle +#1$}}
\atop {\raise1pt\hbox{$\scriptstyle -#2$}}$}}

\def\gsim{\mathrel{\rlap{\lower4pt\hbox{\hskip1pt$\sim$}}
    \raise1pt\hbox{$>$}}}         %greater than or approx. symbol

\def\etal{{\it et al.}}

\def\frac#1#2{{{#1}\over {#2}}}

\def\smallfrac#1#2{\hbox{${{#1}\over {#2}}$}}

\def\GeV{{\rm GeV}}

\def\bq{\bar{q}}
\catcode`@=11 %This allows us to modify plain macros
\renewcommand\section{\@startsection {section}{1}{\z@}
    {-3.5ex plus -1ex minus -.2ex}{2.3ex plus .2ex}{\bf}}
\renewcommand\subsection{\@startsection {subsection}{1}{\z@}
    {-3.5ex plus -1ex minus -.2ex}{2.3ex plus .2ex}{\it}}
\def\slash#1{\mathord{\mathpalette\c@ncel#1}}
 \def\c@ncel#1#2{\ooalign{$\hfil#1\mkern1mu/\hfil$\crcr$#1#2$}}
\def\lsim{\mathrel{\mathpalette\@versim<}}
\def\gsim{\mathrel{\mathpalette\@versim>}}
 \def\@versim#1#2{\lower0.2ex\vbox{\baselineskip\z@skip\lineskip\z@skip
       \lineskiplimit\z@\ialign{$\m@th#1\hfil##$\crcr#2\crcr\sim\crcr}}}
\catcode`@=12 %at signs are no longer letters

\def\PR{{\it Phys.~Rev.~}}
\def\PRL{{\it Phys.~Rev.~Lett.~}}
\def\NP{{\it Nucl.~Phys.~}}
\def\NPBPS{{\it Nucl.~Phys.~B (Proc.~Suppl.)~}}
\def\PL{{\it Phys.~Lett.~}}

\def\ZP{{\it Zeit.~Phys.~}}

\def\APP{{\it Acta.~Phys.~Pol.~}}
\def\vol#1{{\bf #1}}\def\vyp#1#2#3{\vol{#1} (#2) #3}

\def\be{\begin{equation}}
\def\ee{\end{equation}}
\def\bea{\begin{eqnarray}}
\def\eea{\end{eqnarray}}
\def\beq{\begin{equation}}
\def\eeq{\end{equation}}
\def\bq{\begin{quote}}
\def\eq{\end{quote}}
\def\gappeq{\mathrel{\rlap {\raise.5ex\hbox{$>$}} {\lower.5ex\hbox{$\sim$}}}}
\def\lappeq{\mathrel{\rlap{\raise.5ex\hbox{$<$}} {\lower.5ex\hbox{$\sim$}}}}
  \newcommand{\ccaption}[2]{
    \begin{center}
    \parbox{0.85\textwidth}{
      \caption[#1]{\small{\it{#2}}}
      }
    \end{center}
    }
\begin{document}
\parskip 0.3cm

\pagestyle{empty}
\begin{flushright}
{CERN-TH/96-345} \\
{DFTT 77/96} \\
{Edinburgh 96/29}\\
{GEF-TH-22/96}\\
{hep-ph/9701289}
\end{flushright}
%\vspace*{3mm}
\begin{center} {\bf DETERMINATION OF THE BJORKEN SUM AND STRONG COUPLING }\\
               {\bf FROM POLARIZED STRUCTURE FUNCTIONS } \\
\vspace*{0.4cm}  {\bf Guido Altarelli} \\
Theoretical Physics Division, CERN, CH--1211 Geneva 23, Switzerland \\
and Terza Universit\`a di Roma, Rome, Italy \\
\vspace{0.2cm}{\bf
Richard D.Ball\footnote[1]{Royal Society University Research Fellow}} \\
Department of Physics and Astronomy, University of Edinburgh,\\
Mayfield Road, Edinburgh EH9 3JZ, Scotland\\
\vspace{0.2cm}{\bf Stefano Forte} \\
INFN, Sezione di Torino,\\
Via P. Giuria 1, I-10125 Torino, Italy\\
\vspace{0.1cm}{\bf Giovanni Ridolfi} \\
INFN, Sezione di Genova,\\
Via Dodecaneso 33, I-16146 Genova, Italy\\
\end{center}
\vspace*{0.3cm}
\begin{center}
{\bf Abstract}
\end{center}
\noindent
We present a NLO perturbative analysis of all available data on
the polarized structure function
$g_1(x,Q^2)$ with the aim of making a quantitative test of the validity of
the Bjorken sum rule, of measuring $\alpha_s$, and of deriving helicity
fractions. We take particular care over the small $x$ extrapolation,
since it is now known that Regge behaviour is unreliable at perturbative
scales. For fixed $\alpha_s$ we find that if all the most recent data
are included $g_A=1.19\pm0.09$, confirming the Bjorken sum rule at the
$8\%$ level. We further show that the value of $\alpha_s$ is now reasonably
well constrained by scaling violations in the structure function data,
despite the fact that it cannot yet be reliably fixed by the value of
the Bjorken sum: our final result is $\alpha_s(m_Z) = 0.120
\epm{0.010}{0.008}$.
We also confirm earlier indications of a sizeable positive
gluon polarization in the nucleon.

\vfill
\noindent

\begin{flushleft} CERN-TH/96-345 \\ December 1996 \end{flushleft}
\vfill\eject
%\pagestyle{empty}
%\clearpage\mbox{}\clearpage

\setcounter{page}{1} \pagestyle{plain}

%%%%%%%%%%%%%%%%%%%%%%%%%%%%%%%%%%%%%%%%%%%%%%%%%%%%%%%%%%%%%%%%%%%%%%%%%%%%%

Much experimental and theoretical work has been devoted in recent years to
polarized deep inelastic scattering \cite{revs}. Reasonably precise data on
the polarized structure functions of proton
\cite{EMC}-\cite{E143new} and deuteron
\cite{E143new}-\cite{SMCdnew} have been
collected down to values of $x$ near
and below $x=0.01$ for $Q^2>1$~GeV$^2$. Very recently precise results on
the neutron structure function from scattering on {}$^3$He targets
have also become available~\cite{E142,E154}.
The calculation of the kernels for the
perturbative QCD evolution of polarized parton distributions
has recently been completed to next-to-leading order (NLO) \cite{NLO}, thus
reaching the same level of accuracy as in the unpolarized case.
Experience obtained from the small $x$ behaviour of unpolarized
structure functions observed at HERA \cite{revHera} is now
sufficient to indicate at least qualitatively the sort of
behaviour we might expect for polarized structure functions at small $x$.
The purpose of this paper is to take advantage of all this accumulated
knowledge and experience in order to extract from the data on polarized
structure functions the polarized parton densities and their moments for
comparison with theoretical expectations. In particular we will show that
when all the data are included it is now possible to make a reliable
test of the Bjorken sum rule \cite{Bj}. We will also investigate
to what extent one can use the data to make an accurate
determination of $\alpha_s$.

For an experimental verification of the Bjorken sum rule one has to
extract from the data the first moment of the difference of polarized up
and down quark densities at some convenient value of
$Q^2$. Data taken at all kinematically accessible values
of $x$ and $Q^2$, and on all available targets,
contain information relevant for the reconstruction of polarized
parton densities at a given $Q^2$ and ought therefore to be included.
The complete NLO evolution kernels \cite{NLO} can be used to reduce
to the same $Q^2$ data measured at different $Q^2$ for
each $x$.\footnote{In the past this has been done by assuming that the
polarization asymmetries are $Q^2$ independent \cite{EMC,EK93},
but this approximation is questionable given the current precision of the
data.}
Since the evolution equations \cite{AP} for partons at a given $x$ and $Q^2$
depend only on the values of the parton densities at larger values of $x$
and the same $Q^2$, the necessary correction can only be performed
through a general fit to all the data, which yields
a set of polarized parton densities obeying the correct evolution
equations \cite{BFRa,BFRb}. However in order to perform a fit one
must start with a particular ansatz for the parton densities at some reference
$Q_0^2$. Clearly the results of the fit will depend to some extent on
the starting ansatz one adopts, and this dependence will induce an error
in the computed first moments, and in particular in the Bjorken sum. Here
we will devote special attention to this issue.

Once the data are reduced to a common
$Q^2$ for all $x$ values, an extrapolation to unmeasured values at small and
large $x$ is needed in order to obtain the first moment. The extrapolation at
small $x$ is especially
important~\cite{closeroberts}.\footnote{Note that the behaviour at
small $x$ of the input ansatz for the parton densities at $Q_0^2$ is not
relevant for the evolution correction, which only depends on $x$ values
larger than the smallest measured one. On the contrary the integration at
small $x$ that completes a given moment is very much dependent on the
small $x$ behaviour of the input distributions, as we shall see.} In
most of the existing analyses, including those in the experimental papers, it
has been performed by assuming a
simple power behaviour based on Regge theory \cite{Heimann}. This leads
to a rather small contribution to first moments from the small $x$ region,
since the expected extrapolation is at most flat. But two main considerations
are now severely undermining these attempts. First of all the data at
small $x$ for $g_1$ indicate a clear departure from the nearly constant
behaviour expected from Regge theory, pointing to a more singular
behaviour which is moreover different for protons and neutrons. Second,
we have learnt from HERA experiments that the rise of unpolarized
structure functions is also much steeper than that
predicted by Regge theory \cite{revHera}. Although Regge theory seems to work
reasonably well at scales typical of soft hadronic physics, at larger scales
the effects of perturbative QCD evolution are superimposed to give
a rise which gets steeper and steeper as $Q^2$ increases \cite{DeRuj,BF}.
A similar phenomenon is to be expected for the polarized structure
functions \cite{BFRa,prise}, so the results of a naive Regge
extrapolation are not justified if one wants to consider first moments
in the perturbative region: small $x$ contributions to first moments
can be relatively large, especially as $Q^2$ increases.

Here we will discuss alternative extrapolation procedures and the errors
associated with them. Our guiding principles will be the validity of
Regge predictions at low $Q^2$, the buildup with $Q^2$ of the effect of
the QCD evolution and the limits on the size of polarized densities
imposed by the corresponding unpolarized densities. From
these starting points we will estimate the uncertainty in the small $x$
extrapolation, which when combined with the evolution corrections and
the more standard sources of error will allow us to quantify the
extent to which the Bjorken sum rule may be tested using existing data.
In practice we will do this by deriving from the data
the value of $g_A$ and the associated error for an appropriate range
of values of $\alpha_s$.

We will then consider the determination of $\alpha_s$ from the
polarized deep inelastic scattering data. Previous attempts in this direction
\cite{EK} have assumed the validity of the Bjorken sum rule, and used a
value for the Bjorken integral obtained from the first moments given by
the various experimental collaborations, and thus based on naive Regge
extrapolation at small $x$. However when the effects of perturbative
evolution on the small $x$ extrapolation are properly taken
into account, the evaluations of the first moments must be revised:
their errors then turn out to be considerably increased, and the determination
of $\alpha_s$ from the Bjorken sum rule no longer works so well.  However, we
are able to show that a much better determination of $\alpha_s$ may
be obtained if all the available data and not only the Bjorken integral are
used in the analysis: the comparison of the data at small and
large $Q^2$ in the measured range of $x$ then leads to a reasonably
precise measurement of $\alpha_s$.

A similar procedure to that described for the Bjorken sum rule clearly
also applies to the determination of other first moments, and in particular
to the helicity fractions carried by the gluon density and different
flavours of precisely defined quark densities.
Such determinations
using NLO evolution kernels have already been presented in
refs.~\cite{BFRb,polpart}. What is new here, beyond the
inclusion of the recent new experimental results \cite{SMCdnew}-\cite{E154},
is a careful analysis of the dependence of the results on the assumed
form of the input parton densities and thereby on the extrapolation
to small $x$. As a result we confirm the conclusion, first reported in
ref.~\cite{BFRb}, that the data indicate
the presence of a sizeable positive gluon polarization in the proton, as
conjectured several years previously \cite{altarelliross}-\cite{carlitz}.

Finally, we explicitly discuss the possibility of a
direct determination of $\alpha_s$ through the Bjorken sum rule from
the measurement of the isotriplet first moment of $g_1$.
If it were possible to determine this first moment very accurately,
one might then be able to take
advantage \cite{EK} of the fact that radiative corrections to the
nonsinglet first moment are know to NNNLO, that is to order
$\alpha_s^3$. We show that, with current data, even though the central
value of $\alpha_s$ obtained with this procedure is quite compatible
with that of the NLO analysis the experimental error is considerably
larger, due to the uncertainty in the contribution to the isotriplet
first moment from the unmeasured small $x$ region.

\section{Polarized Structure Functions and Partons}

We begin by summarising various results on the relation between structure
functions and polarized parton distributions, and their behaviour at
small $x$, which will be important for the following discussion.

\subsection{Defining Polarized Parton Densities}

The structure function $g_1$ is related to the polarized quark and gluon
distributions by \cite{BFRb}
\beq
g_1(x,Q^2)=\smallfrac{\langle e^2 \rangle}{2} [C_{NS}\otimes \Delta q_{NS}
+C_S\otimes \Delta \Sigma + 2n_fC_g \otimes \Delta g],
\label{1}
\eeq
where $\langle e^2 \rangle=n_f^{-1}\sum_{i=1}^n e^2_i$, $\otimes$ denotes
convolution with respect to $x$, and the nonsinglet and singlet quark
distributions are defined as
\beq
\Delta q_{NS}\equiv \sum_{i=1}^{n_f}
(\smallfrac{e^2_i}{\langle e^2 \rangle} - 1)
(\Delta q_i+\Delta \bar q_i),\qquad
\Delta \Sigma\equiv \sum_{i=1}^{n_f}(\Delta q_i+\Delta \bar q_i),
\label{2}
\eeq
where $\Delta q_i$ and $\Delta \bar q_i$ are the quark and antiquark
distributions
of flavor $i$ and $\Delta g$ is the polarized gluon distribution. The
evolution equations for the polarized parton densities are given by
\bea
\frac{d}{dt}\Delta q_{NS}
&=& \frac{\alpha_s(t)}{2\pi} P^{NS}_{qq} \otimes \Delta q_{NS},\nonumber\\
\frac{d}{dt}\ \left (\matrix{\Delta \Sigma \cr \Delta g}\right)
&=&\frac{\alpha_s(t)}{2\pi}
\left (\matrix{P_{qq}^S & 2n_fP_{qg}^S \cr P_{gq}^S
& P_{gg}^S }\right) \otimes \left (\matrix{\Delta \Sigma \cr \Delta g}\right),
\label{3}
\eea
where $t = \log{Q^2/\Lambda^2}$. The coefficient functions $C$ and the
polarized splitting functions $P$ are now known at LO \cite{AP} and
NLO \cite{NLO}. Moments of coefficient functions and parton
densities are defined as
$f(N) = \int_0^1{dx x^{N-1} f(x)}$ and denoted by
$C(N,\alpha_s)$, $\Delta q_{NS}(N,Q^2)$, $\Delta
\Sigma(N,Q^2)$ and $\Delta g(N,Q^2)$.

As is well known \cite{revs}, the definition of the singlet quark
density $\Delta \Sigma(x,Q^2)$ must be carefully specified, since
its first moment is
maximally scheme dependent. This follows from the fact that, due to the axial
anomaly, $\alpha_s(t)\Delta g(1,Q^2)$ is scale independent at LO.
This implies that even
asymptotically the ambiguity in $\Delta \Sigma(1,Q^2)$ is necessarily of
the same order as its size. For a sensible comparison with the constituent
quark spin one must thus define
$\Delta \Sigma(1,Q^2)$ in such a way that it is scale independent
\cite{altarelliross,AL}: $\Delta\Sigma(1,Q^2)$ =  $\Delta \Sigma(1)$.
This is the definition we will adopt here. We then have
\beq
\Gamma_1(Q^2)\equiv \int_0^1{dx g_1(x,Q^2)}
= \smallfrac{\langle e^2 \rangle}{2}
[C_{NS}(1,\alpha_s(t)) \Delta q_{NS}(1)
+ C_S(1,\alpha_s(t)) a_0(Q^2)],\label{gammaone}
\eeq
with $a_0$ the singlet axial charge: \beq
a_0(Q^2) = \Delta \Sigma(1,Q^2) - n_f \smallfrac{\alpha_s(t)}{2\pi}
\Delta g(1,Q^2).
\label{5}
\eeq
The higher moments of the singlet quark distribution are also scheme dependent,
although in a less dramatic way. Various schemes were
discussed in ref.~\cite{BFRb,FBR} and the dependence of the results of the
analysis on the choice of scheme was studied. Here we do not come back
to this issue but instead simply adopt the AB scheme as defined in
ref.~\cite{BFRb}.
%The component of the theoretical
%error arising from scheme dependence will be taken into account on the
%basis of the results of ref.~\cite{BFRb}.

\subsection{Small $x$ Behaviour}

In view of the need to extrapolate the data to $x = 0$ in order to compute
moments, it is important to summarise the
current understanding of the small $x$
behaviour of structure functions. For the unpolarized singlet quark and
gluon distributions the QCD evolution equations (\ref{3}) lead to the
following asymptotic behaviour at small $x$ \cite{DeRuj,BF}:
\bea
x g &\sim \sigma^{-1/2}e^{2\gamma\sigma-\delta\zeta}
\big(1+\sum_{i=1}^n \epsilon^i\rho^{i+1}\alpha_s^i \big),\nonumber  \\
x \Sigma &\sim \rho^{-1}\sigma^{-1/2}e^{2\gamma\sigma-\delta\zeta}
\big(1+\sum_{i=1}^n \epsilon_f^i\rho^{i+1}\alpha_s^i \big),
\label{smallx}
\eea
where $\xi= \log{x_0/x}$,
$\zeta=\log{\left(\alpha_s(Q_0^2)/\alpha_s(Q^2)\right)}$,
$\sigma=\sqrt{\xi\zeta}$, $\rho=\sqrt{\xi/\zeta}$, and the $\epsilon$
terms indicate corrections from the $n$-th perturbative order
(with $n=1$ corresponding to NLO). It follows that the structure functions
$xF_1$ and $F_2$ rise at small $x$ more and more steeply as $Q^2$ increases,
though, for all finite $n$,
never as steeply as a power of $x$. For all other parton
distributions $f$ ($f = q_{NS}$, $\Delta q_{NS},\Delta\Sigma,\Delta g)$ one has
similarly \cite{prise,BFRa}
\beq
f \sim \sigma^{-1/2}e^{2\gamma_f\sigma-\delta_f\zeta}
\big(1+\hbox{$\sum_{i=1}^n$}
\epsilon_f^i\rho^{2i+1}\alpha_s^i \big).  \label{smallxbis}
\eeq
Thus these distributions are less singular by a factor of $x$ than the
singlet unpolarized distributions eq.~(\ref{smallx}), while the higher
order corrections are more important at small $x$ since the
exponent $i+1$ is replaced by $2i+1$; this is because the leading
small $N$ contributions to the anomalous
dimensions at order $\alpha_s^{i+1}$ are $\left(\alpha_s/(N-1)\right)^i$ in
the unpolarized singlet case, but $N\left(\alpha_s/N^2\right)^i$ for the
nonsinglet and polarized distributions.

The limiting behaviour (\ref{smallx},\ref{smallxbis}) implied by
the evolution equations
(\ref{3}) at finite order in perturbation theory would be spoiled if
the series of higher order powers of $\log {1/x}$ were summed to
all orders to give a
powerlike behaviour in $x$, which would then
overwhelm the leading terms. As is well known~\cite{BFKL}, in
the unpolarized singlet channel one may obtain a
result as singular as $x^{-\lambda}$, with $\lambda\sim 1/2$,
for $x\Sigma$ and $xg$ by
summation of higher order singularities in the Regge limit of
$x\rightarrow 0$ at fixed $Q^2$ (hence fixed $\alpha_s$).
However the meaning and the value of a fixed $\alpha_s$ are quite ambiguous,
and it is not at all necessary a priori that such a singular
behaviour is of relevance in the measured HERA region. In fact the
experimental results from HERA show no evidence at all for this
behaviour \cite{BF,revHera}. In principle the higher order terms could be
more important for the nonsinglet and polarized
distributions due to the $2i+1$ exponent in
eq.~(\ref{smallxbis}) instead of $i+1$ in
eq.~(\ref{smallx}). Indeed summing these `double' logarithmic singularities
\cite{KL,BER} appears to lead to a singular behaviour $f\sim x^{-\lambda}$ with
$\lambda\sim 0.5$ for $q_{NS}$ and $\Delta q_{NS}$, and $\lambda\gappeq 1$ for
the singlet densities (suggesting that the first moment of the singlet part of
$g_1$ could be arbitrarily large). If one were to take these
theoretical predictions seriously, the errors in the small
$x$ extrapolations considered below, particularly in the singlet
channel, would have to be considered only as lower bounds. However the
summation of `double' logarithms is even less well founded theoretically than
the summation of `single' logarithms in the unpolarized singlet channel,
and we believe that at present none of
these results should be interpreted too literally \cite{revHera}.

Another important difference between
the small $x$ behaviour of unpolarized and polarized singlet distributions
is that in the unpolarized case only the gluon anomalous dimension carries
the leading singularity, and consequently the rise in the singlet quark
distribution is driven directly by that of the gluon, while in the
polarized case all the entries in the matrix of singlet anomalous
dimensions are singular, and the polarized singlet quark and gluon
distributions mix. It turns out that the leading eigenvector of small $x$
evolution is then such that the singlet quark and gluon distributions have
opposite sign, which means in practice that the singlet component of $g_1$
is driven negative at small $x$ and large $Q^2$ \cite{BFRa}. Contributions
to first moments of $g_1$ from the small-$x$ tail thus tend to become negative
when $Q^2$ is sufficiently large.

The purely perturbative asymptotic predictions
eqs.~(\ref{smallx},\ref{smallxbis}) only
hold when the input distribution at the starting scale $Q_0^2$ is relatively
nonsingular: if the singularity in the input is stronger than that generated
perturbatively then the input will be essentially preserved by the
perturbative evolution.
The rise at small $x$ will then be largely independent of $Q^2$, rather than
becoming steeper as $Q^2$ increases.
If we take the starting scale in the crossover region
between perturbative and nonperturbative dynamics, we can presumably take the
small $x$ behaviour of the input from Regge theory. For unpolarized
distributions the input to the singlet distributions (given by the pomeron
trajectory) is then relatively flat, and indeed the dominance of the
perturbative behaviour
(\ref{smallx}) is confirmed by $F_2$ data from HERA \cite{revHera,BF}, while
the input to the nonsinglet (given by the $\rho-\omega$
Reggeon trajectory) is singular,
behaving as $x^{-1/2}$, so it is preserved by the evolution and
is consistent with data from NMC and CCFR.
For polarized distributions Regge theory suggests
that the form of the input should be given by the $A_1$ trajectory, and
thus flat or even vanishing, behaving as
$x^{0}$--$x^{0.5}$~\cite{Heimann}.

In the following we will tentatively adopt a physical picture, inspired
by the HERA results, in which we assume the validity of Regge behaviour
at small $x$ in the soft region (i.e. that at some input scale
$Q_0^2\lsim 1$~GeV$^2$ the polarized densities are flat or vanishing)
while at larger $Q^2$ the effect of NLO perturbative
evolution is superimposed (giving a perturbative growth of the form
eq.~(\ref{smallxbis})). We will also allow steeper inputs in the nonsinglet
sector, provided they are consistent with the partonic constraint
$|\Delta q_{NS}(x,Q^2)|\lsim q_{NS}(x,Q^2)$: this limits the effective
small $x$ behaviour of $\Delta q_{NS}$ to a singular growth $\sim x^{-0.5}$.
This picture turns out to be consistent with the data, and gives
a constraint on the allowed growth of $|\Delta q_{NS}(x,Q^2)|$ in
the unmeasured region which in turn limits the possible ambiguity
on the Bjorken sum rule from the small $x$
extrapolation.
%\footnote{Note that it is
%satisfied automatically by inputs which conform to Regge expectations,
%since the growth eq.~(\ref{smallxbis}) is never as fast as a power of $x$.}
Unfortunately in the polarized singlet sector the corresponding
partonic bound is much less restrictive and one has here to rely entirely
on the validity of the NLO approximation and Regge behaviour at a low
enough scale.

\section{Polarized Parton Densities from $g_1$ Data}

We now consider in detail the problem of extracting relevant physical
quantities from the existing data. We devote particular attention to the
study of the dependence of the results on the assumed functional
form of the input parton distributions. For this purpose, we
consider a variety of possible parameterizations of the input,
we evolve these up to the values of $x$ and $Q^2$ where data are available
by solution of the evolution eqs.~(\ref{3}) at NLO, and we determine
the free parameters of the input by a best fit of $g_1(x,Q^2)$
eq.~(\ref{1}) to all the data of
refs.~\cite{SMCp}-\cite{E154} with $Q^2\ge 1$~GeV$^2$.
Experimental data for $g_1$ are obtained from the experimentally
measured asymmetries $A_1$ using a parameterization of the measured
unpolarized structure functions $F_2$ \cite{NMCF2} and
$R$ \cite{SLACR},\footnote{Although for the
E154 data \cite{E154} the preliminary published  values
of $g_1^n$ are used directly since the experimental asymmetries are
not yet available.} consistently
neglecting all higher twist corrections, and, for deuterium
and helium targets, accounting for effects
due to the nuclear wavefunction (but not Fermi motion or shadowing) by
a  simple multiplicative correction~\cite{corrnuc,E142}.
Throughout this section we take
$\alpha_s(m_Z) = 0.118\pm{0.005}$ \cite{alf} and
$a_8=0.579\pm0.025$~\cite{CRb} where
$a_8$ is the SU(3) octet axial charge (in the proton, below charm threshold,
$\Delta q_{NS}(1) \equiv \eta_{NS} = \frac{3}{4}g_A + \frac{1}{4} a_8$).

To begin with we parameterize the initial parton distributions at
$Q_0^2 = 1$~GeV$^2$ according to the conventional form
\beq
\Delta f(x,Q_0^2) = {\cal N}_f \eta_f x^{\alpha_f} (1-x)^{\beta_f}
(1 + \gamma_fx^{\delta_f})
\label{firstclass}
\eeq
where $\Delta f$ denotes $\Delta q_{NS}$, $\Delta \Sigma$ or
$\Delta g$ and ${\cal N}_f$ is a
normalisation factor chosen such that the first moment of
$\Delta f$ is equal to $\eta_f$. The signs
of all parameters are left free, including the overall factors
$\eta_f$ (although the data always choose $\eta_f$ to be positive).
In table 1 we report the results from a fit of this class, denoted as fit A,
which corresponds to:
\beq
\delta_{\Sigma}=\delta_g=1,\qquad\delta_{NS}=0.75\qquad
{\mathrm(fixed)},\qquad\gamma_{\Sigma}=\gamma_g\qquad {\mathrm (fit~A).}
\label{9}
\eeq

It is particularly important to see to what extent the data fix the small $x$
behaviour i.e. the exponents $\alpha_f$ in eq.~(\ref{firstclass}).
Note that the
$g_1$ data on protons and neutrons show a strong rise or fall at small $x$,
while the deuteron data are much flatter. Thus the fitted nonsinglet
quark densities at small $x$ tend to show a
sharp rise while the singlet quarks tend to remain fairly flat. Starting
from a generic input set of densities of the class eq.~(\ref{firstclass})
we can easily end up with $\Delta q_{NS}$ considerably more singular
than $\sim x^{-0.5}$, which we cannot accept since then the nonsinglet
polarized distribution would eventually become larger than the
unpolarized distribution. However a
more careful analysis reveals that there is a strong correlation
between $\alpha_f$ and $\delta_f$. In the measured region one can easily push
$\alpha_{NS}$ closer to zero by decreasing $\delta_{NS}$ from unity without
appreciable changes in the quality of the fit. For example for
$\delta_{NS} = 1,0.75,0.5$ we find $\alpha_{NS} \sim -0.8,-0.5,-0.3$,
respectively, with $\chi^2 \sim 89,89,90$. Thus we
find that the existing data do not much constrain the behaviour of
the nonsinglet at asymptotically small values of $x$: even within
the simple functional form eq.~(\ref{firstclass}) one still has a considerable
flexibility in the asymptotic behaviour as $x\rightarrow 0$. Still,
it appears that, within this class of fits, the
resulting nonsinglet quark density is rather more singular at
$Q_0^2 = 1$~GeV$^2$ than we would expect from naive Regge theory,
suggesting that Regge behaviour can only hold at significantly
lower scales. The choice of parameters eq.~(\ref{firstclass}) corresponds
to a typical case which saturates the partonic
constraint discussed at the end of the previous section.

In order to discuss less singular inputs, we
completely change the functional form of the
input densities (while keeping the initial scale at
$Q_0^2 = 1$~GeV$^2$).
We thus choose an input in which the rise at small $x$ is at most
logarithmic (fit B):
\bea
\Delta \Sigma &=& {\cal N}_{\Sigma}
\eta_{\Sigma}x^{\alpha_{\Sigma}}
\left(\log 1/x\right)^{\beta_{\Sigma}}\nonumber\\
\Delta q_{NS} &=& {\cal N}_{NS} \eta_{NS}
\left[\left(\log 1/x\right)^{\alpha_{NS}} +\gamma_{NS}x
\left(\log1/x\right)^{\beta_{NS}}\right] \qquad {\mathrm (fit~B),}\\
\Delta g &=& {\cal N}_g \eta_g \left[\left(\log 1/x\right)^{\alpha_g}
+\gamma_g x\left(\log 1/x\right)^{\beta_g}\right]\nonumber \label{fitb}
\eea
Note that $\log{1/x}\sim(1-x)$ near $x=1$, so that the
$\gamma$ terms take care of the behaviour as $x\rightarrow 1$. The results
we obtain from this fit are again reported in table 1.
It is remarkable that we obtain a fit that is equally good in the
measured region (as seen from the
$\chi^2$ value) but with a quite different extrapolation in the unmeasured
region. As a consequence, for example, the fitted value of
$g_A$ changes from $g_A=1.17\pm0.05$ in fit A to $g_A=1.23\pm0.07$ in
fit B. The small $x$ behavior of this fit
is weaker than any power, and thus in particular compatible with the Regge
prediction.

Although the logs are reminiscent of QCD evolution the functional form of
fit B might perhaps appear a little ad hoc. It is thus interesting to try to
generate the logarithms in a more physical way, by perturbative evolution.
In this spirit we consider another set of trials, where we start the
QCD evolution at a very small scale, $Q_0^2 = 0.3$~GeV$^2$, and fit a function
of the form eq.~(\ref{firstclass}).
The choice of such a low scale is simply
used as a trick to generate an effective set of distributions at the
 value of $Q^2$ at which we begin to fit the  data, i.e.
$Q^2 = 1$~GeV$^2$ (data with lower $Q^2$ being still discarded),
with the logs piled up in a way entirely consistent with
perturbative evolution. In table 1 we report the results from a fit with
\beq
\gamma_\Sigma=\gamma_g=\gamma_{NS}=0,\qquad
\beta_g=15\qquad {\mathrm (fixed)}\qquad {\mathrm (fit~C).}\label{11}
\eeq
In this class of fits, the large-$x$ behaviour of the gluon distribution
can hardly be determined by the fitting procedure; therefore, we
fixed $\beta_g=15$ at $Q_0^2=0.3$~GeV$^2$, because we checked that this choice
approximately corresponds to a $(1-x)^4$ behaviour of $\Delta g$
at $Q^2$ around 1 GeV$^2$.
Once more the quality of the fit in the measured region is unchanged. Comparing
with the results of fit A, we see that by lowering the initial $Q_0^2$
scale all the exponents in the $x^{\alpha}$ terms have become positive
in qualitative agreement with the idea that naive Regge behaviour is
restored at a sufficiently low scale. Indeed a fit of
comparable quality is obtained if we fix all exponents $\alpha_f$
at $Q_0^2 = 0.3$~GeV$^2$ to the limiting value admitted by Regge theory, i.e.
one half (fit D):
\beq
\alpha_f = 0.5,\qquad
\gamma_{\Sigma}=0,\qquad\delta_g=1,\qquad\delta_{NS}=1,\qquad{\mathrm (fit~D).}
\label{12}
\eeq
The results of this fit are also shown in table 1.
The $\chi^2$ is now slightly worse, but the physical results do not change
much, especially in the nonsinglet sector (for example the central value of
$g_A$ is about  the same in fits C and D). One could presumably optimize
the choice of the initial scale $Q_0^2$ to make the agreement with
Regge theory even better.

In figs.~1a-c we display the best-fit $g_1$ (fit B) for protons, neutrons and
deuterons at the $Q^2$ of the data.
In figs.~2a-c we then compare the best-fit forms of $g_1$ corresponding to
fits A--D at $Q^2=10$~GeV$^2$.
The figures show that while the four fits are reasonably close together in
the measured region ($x\sim0.003-0.03$ up to $x\sim 0.8$), they
become very different in the small $x$ region. Note that at
this  $Q^2$ value $g_1$ has indeed become negative at small $x$ in all
cases.  In fig. 3a-c we display the resulting polarized parton
densities obtained from the fits at the same value of $Q^2$. The behaviour at
small $x$ is quite different in each case. The fits C and D develop a
particularly robust tail at small $x$ in the singlet sector. It is the
large positively polarized gluon that drives $g_1$ negative at small $x$.

In table 2 we report the values obtained by computing the first moments of
$g_1$  by integration of the four fits, both in the measured range of
$x$ and in the whole range at the `average' values of
$Q^2$ quoted by each experiment on protons, deuterons and neutrons. We see
that while the truncated moments are remarkably close to each other
the complete moments show a much wider spread. We also
report the values of the truncated moments obtained by evolving the data to a
common scale by means of the traditional (but
unjustified) assumption that the asymmetries are scale independent
and then summing over the bins, and those
given by the experimental collaborations with their associated total errors.
The latter two values should in principle coincide and only differ because
of details in the way $g_1$ is determined from the measured asymmetries
(such as the use of different
parameterizations of the unpolarized structure function $F_2$, or the
inclusion of some higher twist corrections, as done by some experimental
collaborations).
The effect of the $Q^2$ dependence in the measured region is sizeable
but smaller than the experimental error. Much larger is the indirect effect of
scaling violations on the extrapolation at small $x$ because of the larger
scale dependence at small $x$.

\section{Phenomenological Implications}

We will now discuss the quantitative
consequences that can be derived from the results of the previous section.
In particular we will discuss the
status of the Bjorken sum rule, the polarization of different quark
flavors and of gluons in the proton, and the determination of $\alpha_s$.

\subsection{Testing the Bjorken Sum Rule}

One general way to test the Bjorken sum rule is to determine $g_A$ and the
associated error from fitting the whole set of available data points. The error
has many components to it: the error from the individual fit (based on a given
functional form) due to the experimental errors on each data point, the
ambiguity on the input functional form for the parton densities (which
thus includes the error from the small $x$ extrapolation), the error
due to the uncertainty in the values of $\alpha_s$ and $a_8$
(we take $\alpha_s(m_Z)=0.118\pm0.005$ \cite{alf},
and $a_8=0.579\pm0.025$ \cite{CRb}\footnote{The sensitivity to
$a_8$ is in practice very weak: we have checked that variations of up
to $30\%$ have little effect on our results. Such
deviations from the value of ref.~\cite{CRb} could be caused
by SU(3) violations in hyperon semileptonic decays
\cite{su3}.}), the error due to threshold uncertainties
(estimated by varying the position of the thresholds around $m_q$
by a factor 2), and finally the error related to unknown
higher perturbative orders (estimated by varying
the renormalization and factorization scales $\mu^2_R,\mu^2_F$
by a factor of two in either
direction). The error due to higher twist terms is potentially significant,
because the data at low $Q^2$ (and thus at low $x$) are essential for the
computation of the first moments, but difficult to estimate reliably:
here we estimate it (see sect.~3.4 below)
from the shift in the value of $g_A$ extracted from
the Bjorken sum rule due to the
inclusion of a higher twist correction to it according to
QCD sum rule and renormalon
estimates \cite{ht}.

Based on the fits shown in table 1, and similar ones used to estimate
theoretical errors, we find
\beq
g_A = 1.19\pm0.05{(\mathrm exp)}\pm 0.07{(\mathrm th)}
=1.19\pm0.09, \label{gafit}
\eeq
where the central value is obtained as the average between the maximum and 
minimum values of table 1, and
the various contributions to the theoretical error are listed in
table~3. The fitted value is to be compared with the direct measurement
$g_A = 1.257\pm0.003$ \cite{PDG} from $\beta$-decay. Thus we find that
the Bjorken sum rule is confirmed to within one standard deviation but
still with an accuracy of only about 8\%. For the chosen
value of $\alpha_s$ the fitted value of $g_A$ is below the experimental
value; it would be increased by using a larger value of $\alpha_s$, or
perhaps by the inclusion of additional corrections beyond NLO.

\subsection{Singlet First Moments}

Similarly in the singlet sector one obtains from the data values for
$\eta_q=\Delta \Sigma(1)$ (the conserved polarized singlet quark
density), for $\eta_g=\Delta g(1,1\,{\mathrm GeV}^2)$
(the first moment of the polarized
gluon density evaluated at $Q^2=1$~GeV$^2$), and for $a_0(10\,{\mathrm GeV}^2)$
(the non
conserved singlet axial charge defined implicitly from the singlet part of
$g_1$ by eq.~(\ref{gammaone})).
This latter quantity approaches a finite limit at
infinite $Q^2$ because the corresponding anomalous dimension starts at
two loops, and within the present accuracy $a_0(10\,{\mathrm GeV}^2)$ is
equivalent to $a_0(\infty)$. The values for these three
quantities as obtained from our representative fits are reported in table 1. We
then studied in detail, following ref.~\cite{BFRb}, the theoretical
errors from the various different sources: results for these
are listed in table~3. The error due to higher twists is not included
since it cannot be reliably estimated but, on the basis of the corresponding
error on $g_A$, it is expected to scarcely affect the total errors which
in the singlet sector are rather large (for reasons discussed in detail
in refs.~\cite{BFRa,BFRb}).
  In comparison to ref.~\cite{BFRb} we have
a more reliable determination of the error related to the fitting
procedure, and a somewhat stronger sensitivity to higher order
corrections. We then find \bea
\Delta \Sigma (1) &=& 0.45\pm0.04~{\mathrm (exp)}\pm0.08~{\mathrm (th)}
= 0.45\pm0.09, \nonumber\\
\Delta g(1,1\,{\mathrm GeV}^2)&=&
1.6\pm0.4~{\mathrm (exp)}\pm0.8~{\mathrm (th)} = 1.6\pm0.9,\label{14}\\
a_0(\infty)&=&0.10\pm0.05~{\mathrm (exp)}\epm{0.17}{0.10}~{\mathrm (th)}
= 0.10\epm{0.17}{0.11},\nonumber
\eea
in excellent agreement with
the corresponding results of ref.~\cite{BFRb}. The physical implications
are thus the same. The parameter $a_0(\infty)$ measures the
degree of `spin crisis': the singlet axial charge of the nucleon is still
compatible with zero as it was at
the beginning of the story \cite{EMC}. Note that with the naive Regge
extrapolation at small $x$ the experimental result for the axial singlet
charge would be significantly larger, with a much smaller error: for example
in ref.~\cite{AlRi} a value $a_0(\infty)=0.33\pm0.04$ was quoted.
There is also some evidence (around two standard deviations) for a
positive gluon
polarization in the nucleon (increasing with $Q^2$ as $1/\alpha_s(Q^2)$).
The amount of gluon polarization is large enough to allow the first moment
$\Delta \Sigma (1)$ of the conserved singlet quark density to be within one
standard deviation of $a_8 \sim 0.58$,
which in the absence of all SU(3) and chiral symmetry breaking effects, could
be identified with the constituent spin fraction \cite{EJ}. This
can be seen as a direct
confirmation of the physical explanation of the `spin crisis', advocated in
refs.~\cite{altarelliross}-\cite{carlitz}, as due to the anomaly and
well described in terms of the QCD parton model (for more general
possibilities, see refs.~\cite{instantons}).

\subsection{Determination of $\alpha_s$}

The above discussion on $g_A$ makes it clear that the determination
of $\alpha_s$
from the Bjorken integral is adversely affected by the increased ambiguity
from the small $x$ extrapolation that follows from the
demise of the naive Regge behaviour at small $x$. Fortunately we find that
$\alpha_s$ can be determined directly from the available data without
extrapolation in the small $x$ region
if the totality of the data is taken into account and not just the Bjorken
integral. The value of $\alpha_s$ is then determined by the strong scaling
violations needed to accommodate the
difference between the data at small $Q^2$ from the SLAC experiments and
those at larger $Q^2$ from the SMC in the common
range of $x$. While $\Delta g(1,1\,{\mathrm GeV}^2)$
is mainly fixed by the proton data,
$\alpha_s$ is determined by the difference between proton and neutron.
To show this we repeated the fits A-D but fixing $g_A$ to its
experimental value and instead fitting $\alpha_s$ (in this case,
fit A was performed with $\beta_g$ fixed at the value $\beta_g=4$).
In all cases the central value was found to be close to $\alpha_s(m_Z)=0.120$.
Since the different
fits differ considerably in the unmeasured region, this shows that
it is the behaviour in the measured region that matters. In addition, the
Bjorken integral is appreciably different in the different fits
A-D when $\alpha_s$ is kept fixed
(and the value of $\Delta g(1,1\,{\mathrm GeV}^2)$ even more so)
but this difference does not affect the fitted $\alpha_s$ very much.
Furthermore, the resolution of the discrepancy between the fitted
value of $g_A$ eq.~(\ref{gafit}) and its
experimental value would require a much larger increase of $\alpha_s$
if only the Bjorken integral were relevant for fixing $\alpha_s$. These results
show that $\alpha_s$ is much better constrained by the overall
pattern of scaling violations than by the Bjorken integral alone.

The theoretical uncertainties that affect this determination of $\alpha_s$
are listed in table~3. The main source
of uncertainty originates from higher order and higher twist contributions.
Higher order corrections are  estimated by varying
the renormalization and factorization scales $\mu^2_R$, $\mu^2_F$ around
$Q^2$. The values of table~3 are obtained varying $\mu^2_R$ and $\mu^2_F$
between $Q^2/2$ and $4Q^2$.
A wider scale variation at the lower edge is made impossible, without
leaving the perturbative domain, by
the presence of data with values of $Q^2$ close to 1~GeV$^2$.
The higher twist error is again determined from
the corresponding shift in the value of $\alpha_s$ extracted from
the Bjorken sum rule (sect.~3.4 below).
We  can cross-check this estimate of higher twist terms
by repeating the fit while excluding all data with $Q_0^2<2$~GeV$^2$.
We find that
the value of $\alpha_s$ does not change, thus suggesting that higher
twist effects are indeed quite small.

We thus obtain finally
\beq
\alpha_s(m_Z) = 0.120\epm{0.004}{0.005}{\mathrm (exp)}
\epm {0.009}{0.006}{\mathrm (th)}
= 0.120\epm{0.010}{0.008}.\\
\label{alphafit}
\eeq
This reasonably good determination of $\alpha_s(m_Z)$ could still be
improved with better data:
it is important to notice that without the very recent neutron data \cite{E154}
the experimental error would be twice as large.
We see no reason why it could not be as good as the determination
from unpolarized data if more data were added and the experimental
errors consequently reduced. However the theoretical error is already the
dominant one; it could also be reduced by more data because it is made
particularly large by the small $Q^2$ values of the neutron measurements.

\subsection{Determining $\alpha_s$ from the Bjorken sum rule}

Even though the full set of anomalous dimensions and coefficient functions
are known at NLO, for the nonsinglet first moment the anomalous dimension
vanishes and the coefficient function is known to order $\alpha_s^3$, i.e.
at NNNLO~\cite{NNNLO}.
One would like to take advantage of this situation by
directly determining $\alpha_s$ from the Bjorken sum rule \cite{EK}.
This however requires first a determination of
the Bjorken integral from the data. Since the latter can only be accomplished
at NLO the procedure is only advantageous to the extent that this extraction
can be done in a way that minimizes the necessary theoretical input.

This can be done following a procedure close to that
used in the experimental papers:  first one computes
directly from the data the contribution
to the first moment from the experimentally accessible $x$ range
(applying a NLO correction to account for evolution of the data
at a common scale for each experiment),
then one adds to it an extrapolation over the
unobserved $x$-range. The resulting values of the first moments
are finally analysed using the NNNLO result for the Bjorken
sum rule in order to extract $\alpha_s$.
The procedure can be  justified if the
$Q^2$ span of the data around the average value
in each experiment is small in comparison to the difference in
average $Q^2$ between experiments.
This is indeed the case for most of the data \cite{SMCp}-\cite{E154},
with the exception of the SMC data at very large or very small $x$,
which however give a relatively small contribution to the first moments.
Although it is generally inconsistent to correct the data by using
scaling violations at NLO and then to pretend one has a NNNLO
accuracy for first moments, in practice this might not be a
cause for real concern since the corrections to the data on the
truncated first moments from the NLO scaling
violations are not large in comparison with the experimental errors and
those from the extrapolation in the unmeasured region (see table 2).

We thus first determine for each experiment the value of
the first moment of $g_1$ in the measured region, evaluated at the average
scale $\langle Q^2\rangle $
quoted for that particular experiment, by simply summing the
experimental determinations $g_1^e$ over the bins. Corrected values
of $g_1$ in order  to account
for evolution at a common scale are determined according to~\cite{SMCdnew}
\bea
&&g_1(x,\langle Q^2\rangle)\approx g_1^e(x, Q^2)+\Delta(x,Q^2,\langle
Q^2\rangle ),\\
&&\Delta(x,Q^2,\langle Q^2\rangle )\equiv
g_1^f(x, \langle Q^2\rangle)-g_1^f(x, Q^2),
\label{15b}
\eea
where $g_1^f(x,Q^2)$ is a best-fit,
which we take from our sets of fits A-D. The fits with
$g_A$ left as a free parameter must be used here, otherwise the determination
of the nonsinglet first moment would be circular.
Because all these fits
give very similar results in the measured region (compare table~2),
we can assume the error in the evolution procedure to be negligible
and take the error in this contribution to the moment to be that
given by the experimental collaborations.
We  then complete the moments in the unmeasured region by
again using the set of fits A--D, which differ widely in the small $x$ region,
but are all consistent both with the experimental data and well understood
theoretical principles. Specifically, we take (for each experiment)
the average of the difference between the full and truncated integrals
for the four fits  as an estimate of the extrapolation. The associated
uncertainty is then taken to be given by their spread
as one varies the functional
form of the fit, as well as renormalization and factorization
scales, and  the value of $\alpha_s$ (variations of the thresholds
and the value of $a_8$ have no significant effect).
The measured and extrapolated first moments are then added and the
respective errors added in quadrature (see table 4). This is
then  the best
estimate of the experimental determination of the first moments and their
associated errors.

In order to determine the isotriplet combinations relevant for the
Bjorken sum rule we have repeated the same analysis directly at the
level of the isotriplet combinations of first moments, which we can determine
separately at 10~GeV$^2$ for the SMC experiment, and at
3~GeV$^2$ for all of the SLAC experiments (by averaging the neutron data).
The error due to the extrapolation is then only sensitive to the uncertainty
in the nonsinglet contribution to $g_1$ at small $x$ and thus much smaller
than the sum in quadrature of the errors in individual first moments:
in practice only the variation of the functional form of the fit
contributes, because the nonsinglet extrapolation is essentially insensitive
to the value of $\alpha_s$ and the choice of renormalization and
factorization scales.
The result, listed in table~4, shows that whereas for the SMC data
the dominant uncertainty is in the measured region, for the SLAC
data the uncertainty in the extrapolation is comparable to it.

{}~From the isotriplet first moments of table~4 we can
then determine $\alpha_s(m_Z)$ by fixing $g_A$ to its value
$g_A=1.2573\pm0.0028$,
eq.~(\ref{gafit}). We find
\beq
\alpha_s(m_Z) = \cases{0.111\epm{0.035}{0.111}~{\mathrm (exp)}& SMC only,\cr
                       0.118\epm{0.010}{0.026}~{\mathrm (exp)}& SLAC only,\cr
                       0.118\epm{0.010}{0.024}~{\mathrm (exp)}& all data,\cr}
\label{astwostep}
\eeq
where for the SMC only case the lower
error should be understood as meaning that the uncertainty is
100\% or more. The error in eq.~(\ref{astwostep})
is the purely  experimental one due to the error
on the isotriplet first moment of table 4 . This error is now so large
that it dominates any theoretical errors.
For instance, we can estimate a  higher twist error by adding to
the Bjorken integral a twist four term, with a coefficient estimated
from sum rule and renormalon
calculations \cite{ht}. The value of $\alpha_s$ of eq.~(\ref{astwostep})
is then reduced
by about $0.004$. We used this as an  estimate of the error from
higher twist terms in the determination of $\alpha_s$ eq.~(\ref{alphafit})
(see table~3).

As a cross-check, we can instead fix $\alpha_s$ to its known value
$\alpha_s(m_Z)=0.118\pm0.005$ and
extract $g_A$ from the sum rule: we then get $g_A=1.27\pm0.11\pm0.05\pm0.03$
where the first error is experimental,
the second is due to the error on $\alpha_s$, and the third is the
higher twist error, estimated as above (and used in table 3).
These errors are comparable
to those found by the direct determination of $g_A$ eq.~(\ref{gafit}),
but the experimental error is somewhat larger because only a subset
of the information contained in the data is now being used.

We conclude that the result for $\alpha_s$
obtained from this procedure
is compatible with that from the direct determination
eq.~(\ref{alphafit}), but that the error is now considerably larger, in
agreement with our previous observation that
$\alpha_s$ is better constrained by the data in the measured
range than by the Bjorken integral alone.
The value of $\alpha_s(m_Z)$ is very close to that found in
ref.~\cite{EK} but the error is larger because of the error
from the small $x$ extrapolation which we no longer assume to follow
naive Regge expectations at $\langle Q^2\rangle$.
A direct determination of $\alpha_s$ from the
Bjorken sum rule could however be competitive if data with a wider kinematic
coverage in $x$ were available.

\section{Conclusions}

We have performed a global analysis of all data on the polarized structure
function $g_1$ for proton, deuteron and neutron targets, including
data which have only recently become available,
using NLO perturbative QCD. We took
care to properly estimate the effects of perturbative evolution and the
uncertainties in the small $x$ extrapolation when computing first moments.
We showed that the data confirm the Bjorken sum rule to within one standard
deviation at the $8\%$ level, and indicate a gluon polarization in the nucleon
which is non-zero at the level of two standard deviations.
We showed
further that the data now provide a reasonably accurate determination of
$\alpha_s$, consistent with the global average. However we also showed that
the Bjorken integral is unfortunately not yet sufficiently well determined
to admit a competitive determination of $\alpha_s$ using the Bjorken sum rule.

Our determination of $\alpha_s$ could be improved by
more accurate data on polarization asymmetries in the fixed target region,
or by an independent (semi-inclusive) determination of the polarized
gluon distribution. However the theoretical error from higher order
corrections is already the dominant one. The error in our determination
of first moments is instead dominated by uncertainties in the small $x$
region, and we suspect that these uncertainties will only be significantly
reduced by the measurement of $g_1$ at lower $x$ and higher $Q^2$, which would
be possible at a polarized colliding beam experiment at HERA \cite{HERAS}.

\bigskip
\noindent
{\bf Acknowledgement}: We thank P.~Bosted  and 
A.~Deshpande for useful information on the experimental data.

\vfill\eject

\vfill\eject

\begin{center}
\begin{table}
\vspace*{0.5cm}
\begin{center}
\begin{tabular} {|l|l|l|l|l|}
\hline Parameters&A&B&C&D\\
\hline
d.o.f. & $114-11$ & $114-11$ & $114-8$ & $114-8$\\
$Q_0^2/\GeV^2$ & 1 & 1 & 0.3 & 0.3\\
\hline
$\eta_{\Sigma}$ & $0.408\pm0.041$ & $0.410\pm0.039$ &
$0.422\pm0.026$ &$0.492\pm0.036$\\
$\alpha_{\Sigma}$ & $0.741\pm0.353$ & $1.710\pm0.416$ & $2.600\pm0.964$ &
$0.5$~~~(fixed)\\
$\beta_{\Sigma}$ & $3.105\pm1.049$ & $2.735\pm0.466$ & $3.359\pm1.210$ &
$1.039\pm0.241$\\
$\gamma_{\Sigma}$ & $0.185\pm2.496$ & $0$~~~(fixed) & $0$~~~(fixed) &
$0$~~~(fixed)\\
$\eta_g$ & $1.068\pm0.403$ & $1.032\pm0.330$ &
$0.479\pm0.095$ &
$0.650\pm0.104$\\
$\alpha_g$ & $-0.597\pm0.286$ & $2.970\pm0.611$ & $0.217\pm0.319$ &
$0.5$~~~(fixed)\\
$\beta_g$ & $0.831\pm2.322$ & $1.286\pm0.895$ & $15$~~~(fixed) &
$13.19\pm9.57$\\
$\gamma_g$ & $0.185\pm2.496$ & $21.2\pm22.5$ & $0$~~~(fixed) &
$-0.548\pm9.139$\\
$g_A$ & $1.168\pm0.052$ & $1.234\pm0.066$ & $1.146\pm0.038$ &
$1.141\pm0.036$\\
$a_8$ & $0.579$~~~(fixed) & $0.579$~~~(fixed) & $0.579$~~~(fixed) &
$0.579$~~~(fixed)\\
$\alpha_{NS}$ & $-0.537\pm0.057$ & $1.656\pm0.166$ & $0.765\pm0.228$ &
$0.5$~~~(fixed)\\
$\beta_{NS}$ & $2.503\pm0.274$ & $5.320\pm0.251$ & $2.087\pm0.448$ &
$2.622\pm0.410$\\
$\gamma_{NS}$ & $17.46\pm8.43$ & $-0.229\pm0.105$ & $0$~~~(fixed) &
$5.024\pm4.647$\\
\hline
$\chi^2$ & 83.7 & 83.0 & 83.8 & 90.9\\
$\chi^2$/d.o.f. & $0.813$ & $0.806$ & $0.790$ & $0.858$\\
\hline
$\Delta g(1,1 \GeV^2)$ & $1.07\pm0.40$ & $1.03\pm0.33$&
$1.61\pm0.32$ & $2.08\pm0.34$\\
$a_0(10 \GeV^2)$ & $0.15\pm0.07$  & $0.16\pm0.05$ & $0.05\pm0.04$ &
$0.02\pm0.04$\\
\hline
\end{tabular}
\end{center}
\caption{Results of fits A--D described in the text}
\end{table}
\end{center}

\begin{center}
\begin{table}
\vspace*{0.5cm}
\begin{center}
\begin{tabular} {|l|l|l|l|l|l|l|}
\hline
 & SMC~:~p &E143~:~p & SMC~:~d & E143~:~d & E142~:~n & E154~:~n\\
\hline
$\langle Q^2\rangle /\GeV^2$ & 10 & 3 & 10 & 3 & 2 & 5 \\
\hline
Meas.~Range:~Exp. & 0.1310 & 0.1200 & 0.0379 &
0.0400 & $-0.0280$ & $-0.0370$\\
Exp.~Error &$\pm0.0156$ & $\pm0.0089$ & $\pm0.0079$ &
$\pm0.0050$ & $\pm0.0085$ & $\pm0.0108$\\
\hline
$A_1$ ind. $Q^2$ & 0.1350 & 0.1059& 0.0442& 0.0398& $-0.0293$ & $-0.0364$\\
\hline
Meas.~Range:~A & 0.1256 & 0.1070 & 0.0430 & 0.0378 & $-0.0328$ & $-0.0338$\\
Meas.~Range:~B & 0.1268 & 0.1061 & 0.0425 & 0.0373 & $-0.0326$ & $-0.0342$\\
Meas.~Range:~C & 0.1280 & 0.1084 & 0.0462 & 0.0401 & $-0.0289$ & $-0.0316$\\
Meas.~Range:~D & 0.1308 & 0.1083 & 0.0492 & 0.0402 & $-0.0306$ & $-0.0308$\\
\hline
Full~Range:~A & 0.1171 & 0.1137 & 0.0275 & 0.0265 & $-0.0600$ & $-0.0614$\\
Full~Range:~B & 0.1232 & 0.1197 & 0.0286 & 0.0276 & $-0.0637$ & $-0.0654$\\
Full~Range:~C & 0.1039 & 0.0999 & 0.0160 & 0.0144 & $-0.0707$ & $-0.0716$\\
Full~Range:~D & 0.0990 & 0.0946 & 0.0115 & 0.0094 & $-0.0754$ & $-0.0759$\\
\hline
\end{tabular}
\end{center}
\caption{Determination of the first moment $\Gamma_1(\langle Q^2\rangle)$
eq.~(\ref{gammaone}).
For each experiment we display the average value of $Q^2$ and the
contribution to the first moments from the measured range of $x$,
as given, first, by the experimental collaborations, with the
corresponding total (statistical and systematic) error,
then by summing over experimental bins while evolving the data
assuming  scale independent asymmetries, and finally as obtained from
integration of
the fits A--D. In the last four rows the complete first
moments obtained from the fits A--D are shown.}
\end{table}
\end{center}

\begin{center}
\begin{table}
\vspace*{0.5cm}
\begin{center}
\begin{tabular} {|l|l|l|l|l|l|l|}
\hline  & $g_A$ &$\Delta\Sigma$& $\Delta g$
&$ a_0$ & $\alpha_s$\\
\hline
experimental  & \apm0.05 & \apm0.04 & \apm0.4 & \apm0.05 &
\epm{0.004}{0.005} \\
\hline
fitting       & \apm0.05 & \apm0.05 & \apm0.5 & \apm0.07 & \apm0.001 \\
$\alpha_s$ \& $a_8$ & \apm0.03 & \apm0.01 & \apm0.2 &
\apm0.02 & \apm0.000 \\
thresholds    & \apm0.02 & \apm0.05 & \apm0.1 & \apm0.01 & \apm0.003 \\
higher orders   & \apm0.03 & \apm0.04 & \apm0.6 &
\epm{0.15}{0.07} & \epm{0.007}{0.004} \\
higher twists       & \apm0.03 &  -   &  -  &  -  & \apm0.004 \\
\hline
theoretical         & \apm0.07 & \apm0.08 & \apm0.8 &
\epm{0.17}{0.010} &\epm{0.009}{0.006} \\
\hline
\end{tabular}
\end{center}
\caption{Contributions to the errors in the determination of the quantities
$g_A$, $\Delta\Sigma(1)$, $\Delta g(1,1\GeV^2)$, $ a_0(\infty)$ and
$\alpha_s(m_Z)$ from the fits described in the text.
}
%\caption{sempre caro mi fu}
\end{table}
\end{center}

\begin{center}
\begin{table}
\vspace*{0.5cm}
\begin{center}
\begin{tabular} {|l|l|l|l|l|l|l|}
\hline  & $\langle Q^2\rangle$ &Meas. Range &Full~Range \\
\hline
{}~SMC~:~p & 10 & $0.133\pm 0.016$ & $0.116\pm 0.022$\\
{}~SMC~:~d & 10 & $0.045\pm 0.008$ & $0.021\pm 0.016$\\
{}~SMC~:~I=1 & 10 & $0.176\pm 0.036$ & $0.191\pm 0.037$\\
\hline
{}~E143~:~p &  ~3  & $0.108\pm 0.009$& $0.107\pm 0.017$\\
{}~E143~:~d &  ~3  & $0.040\pm 0.005$& $0.021\pm 0.014$\\
{}~E142~:~n &  ~2   & $-0.032\pm 0.009$& $-0.068\pm 0.015$ \\
{}~E154~:~n &  ~5   & $-0.034\pm 0.011$& $-0.070\pm 0.015$ \\
SLAC~:~I=1 &  ~3   & $0.139\pm 0.013$& $0.177\pm 0.018$ \\
\hline
\end{tabular}
\end{center}
\caption{Best estimates and errors for the truncated and the complete first
moments for the target and
the average $Q^2$ that correspond to each experiment. The isotriplet
first moments for the SMC experiment, and
for all the SLAC experiments combined are also given.}
\end{table}
\end{center}
\vfill
\eject
%%%%%%%%%%%%%%%%%%%%%%%%%%%%%%%%%%%%%%%%%%%%%%%%%%%%%%%%%%%%%%%%%%%%%%%
\begin{figure}[ptbh]
  \begin{center}
    \mbox{
      \epsfig{file=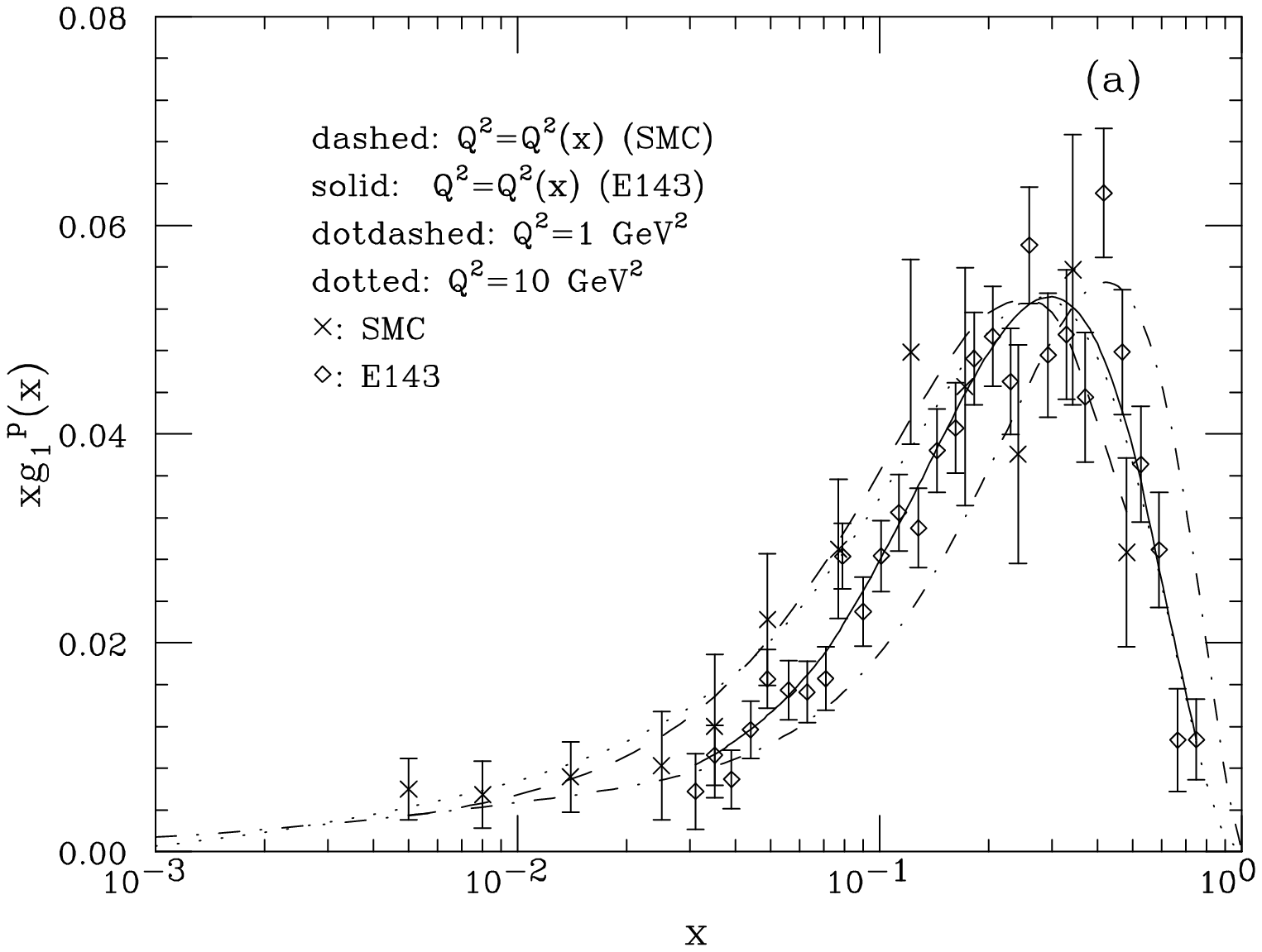,width=0.70\textwidth}
      }
  \end{center}
\end{figure}
\begin{figure}[ptbh]
  \begin{center}
    \mbox{
      \epsfig{file=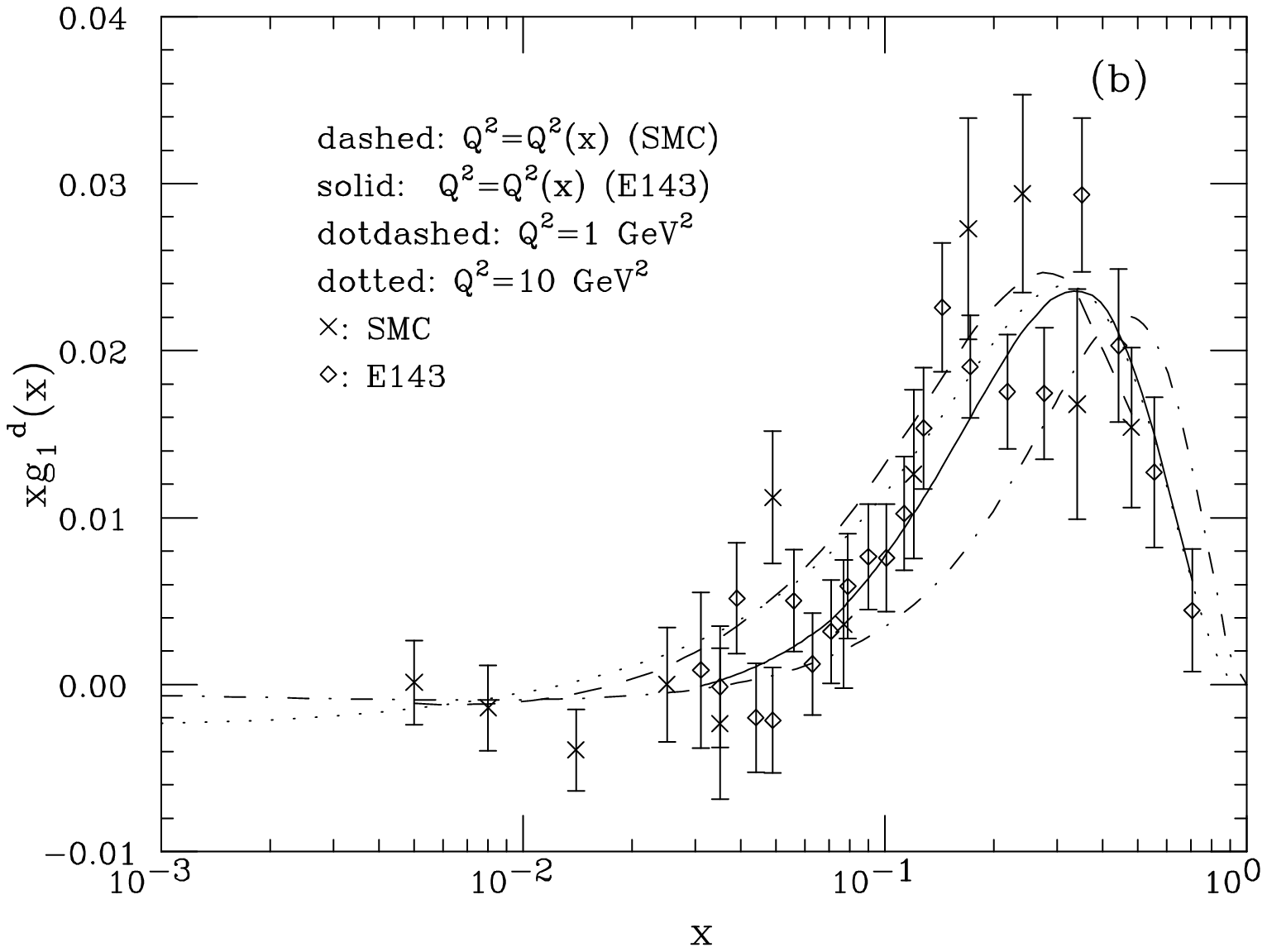,width=0.70\textwidth}
      }
  \end{center}
\end{figure}
\begin{figure}[ptbh]
  \begin{center}
    \mbox{
      \epsfig{file=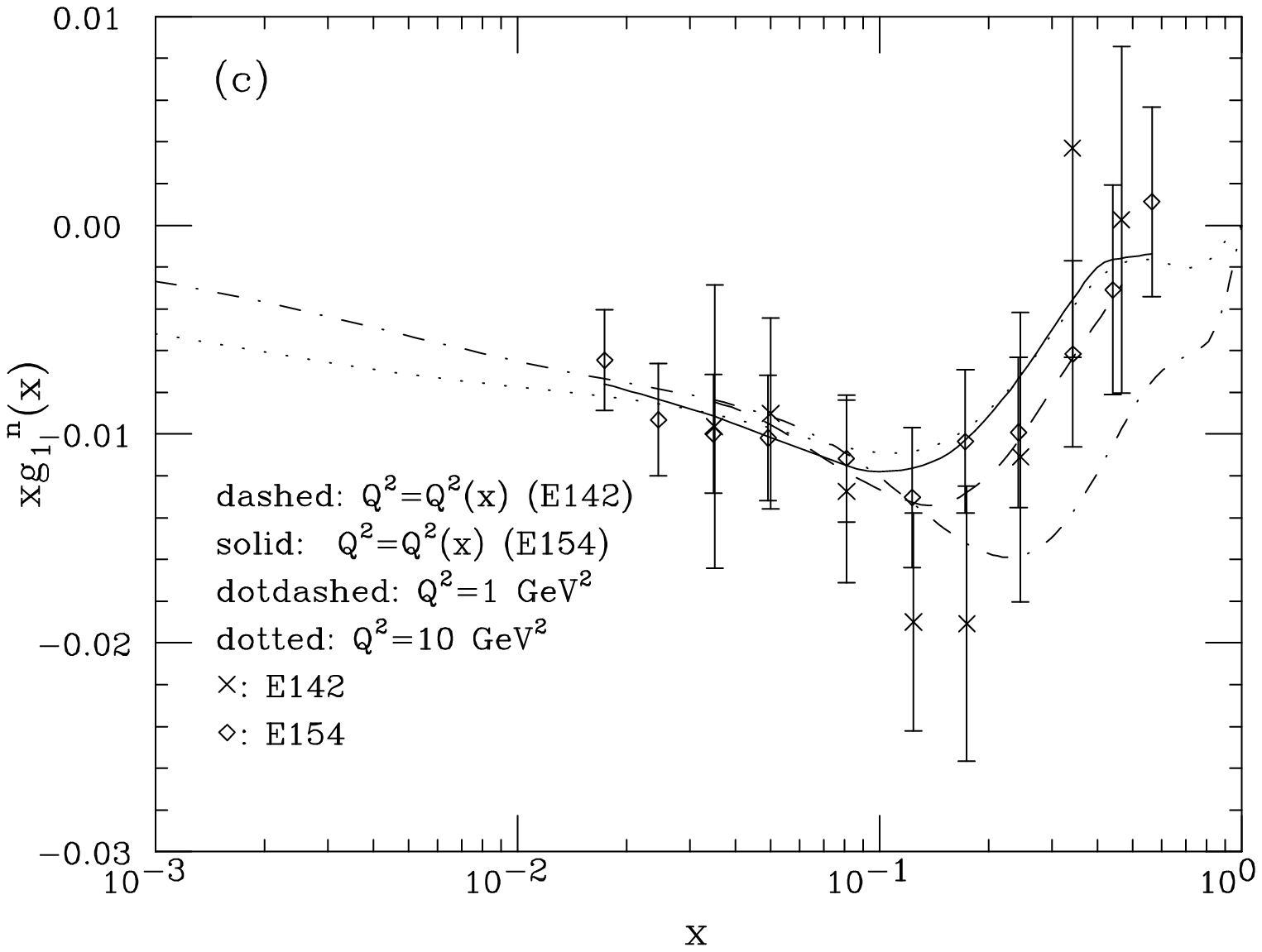,width=0.70\textwidth}
      }
  \ccaption{}{
Plots of $x g_1(x,Q^2)$ for fit B
for(a)  proton (b) deuterium (c) and neutron  targets. The curves
correspond to
$Q^2=1$~GeV$^2$ (dotdashed), $Q^2=10$~GeV$^2$ (dotted),
and $Q^2(x)$ of the various experiments: E143 (solid) and SMC (dashed)
for figs.~a-b and  E142 (solid)
and E145 (dashed) for fig.~c. The data points with total errors are also shown.
}
  \end{center}
\end{figure}
%%%%%%%%%%%%%%%%%%%%%%%%%%%%%%%%%%%%%%%%%%%%%%%%%%%%%%%%%%%%%%%%%%%%%%%
%%%% end fig 1
%%%%%%%%%%%%%%%%%%%%%%%%%%%%%%%%%%%%%%%%%%%%%%%%%%%%%%%%%%%%%%%%%%%%%%%
\begin{figure}[ptbh]
  \begin{center}
    \mbox{
      \epsfig{file=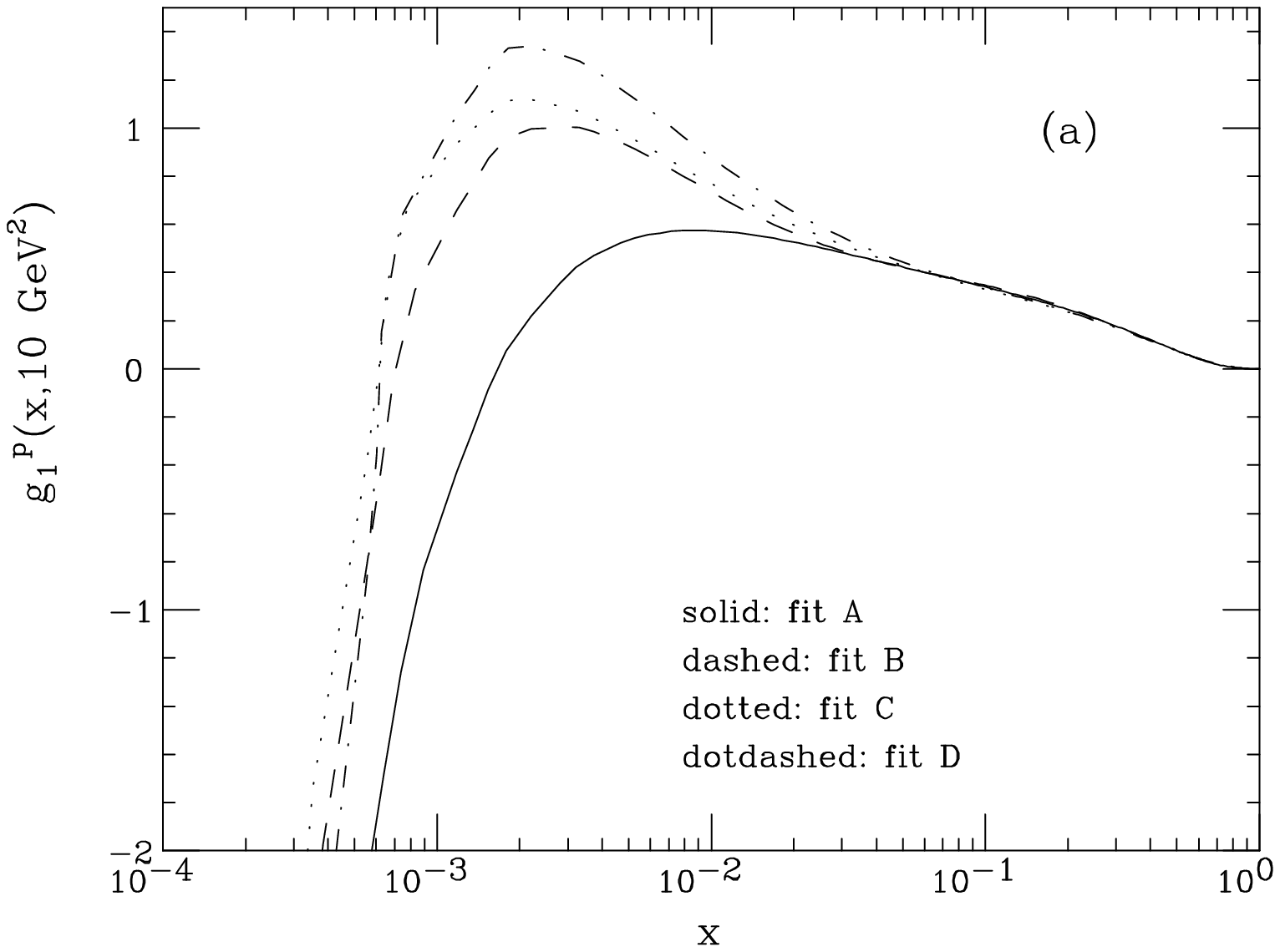,width=0.70\textwidth}
      }
  \end{center}
\end{figure}
\begin{figure}[ptbh]
  \begin{center}
    \mbox{
      \epsfig{file=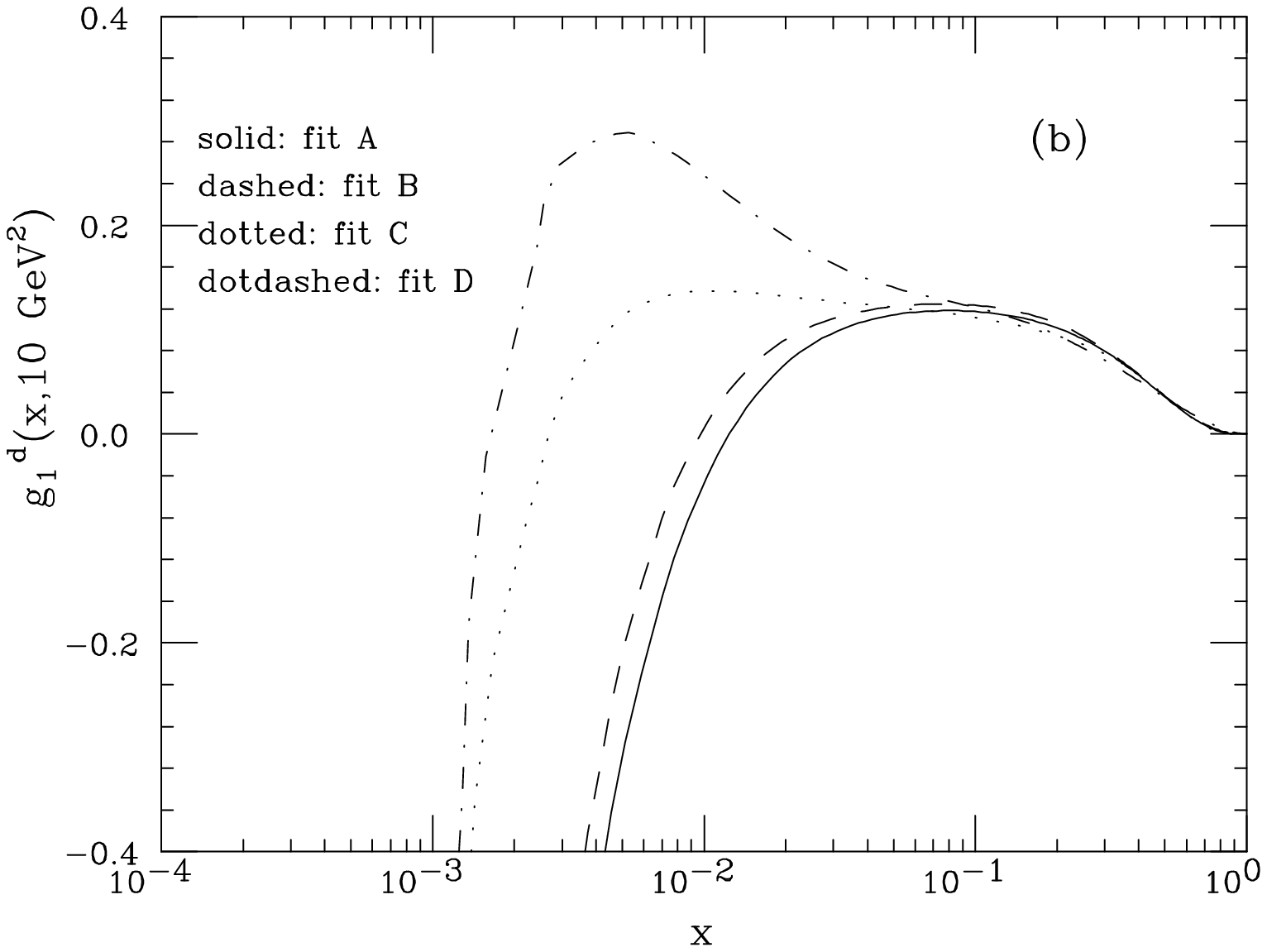,width=0.70\textwidth}
      }
  \end{center}
\end{figure}
\begin{figure}[ptbh]
  \begin{center}
    \mbox{
      \epsfig{file=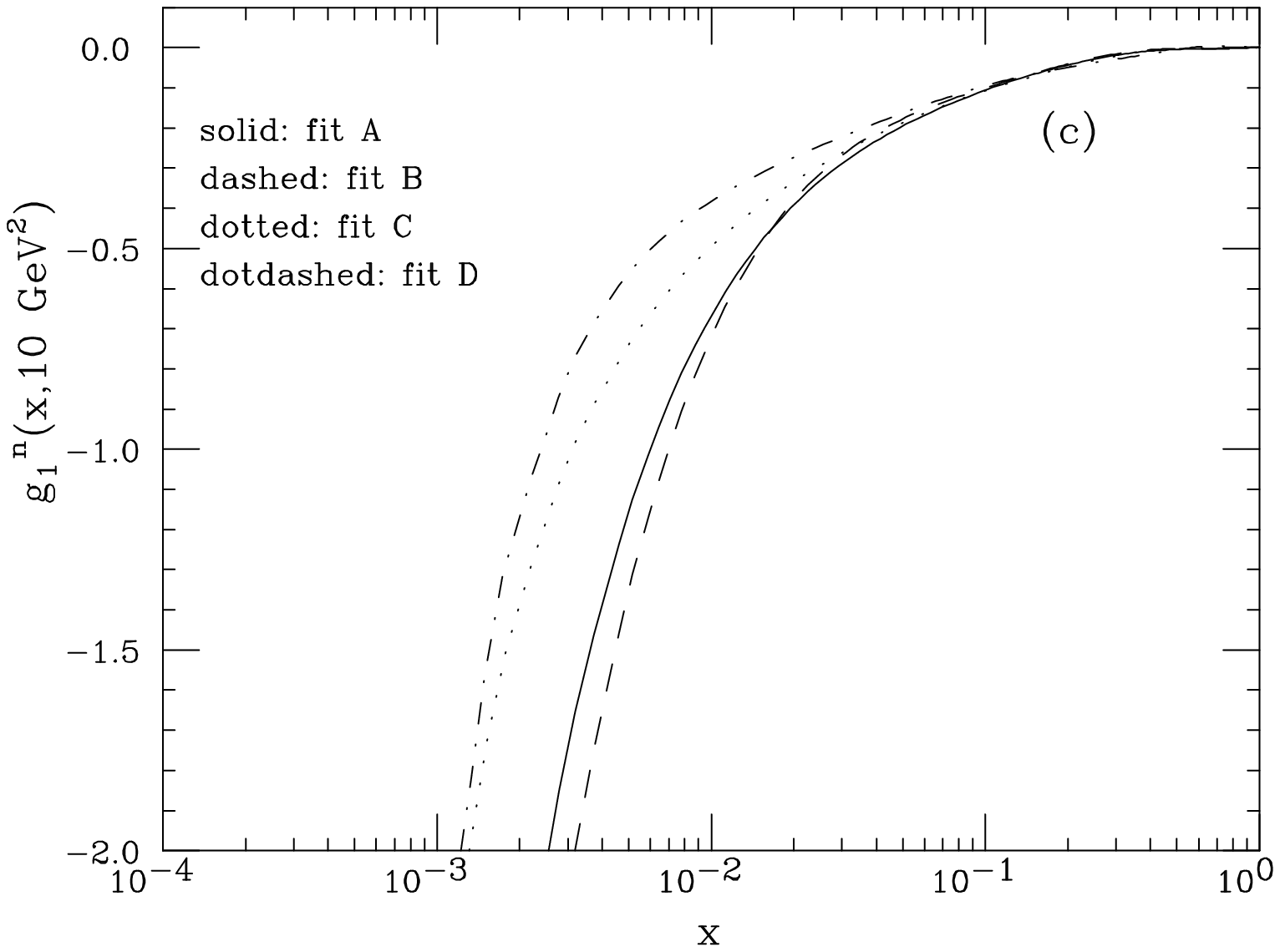,width=0.70\textwidth}
      }
  \ccaption{}{
Plots of $g_1(x,Q^2)$
for (a)  proton (b) deuterium (c) and neutron  targets
 for fits A--D at
$Q^2=10$~GeV$^2$.
}
  \end{center}
\end{figure}
\vfill
\eject
%%%%%%%%%%%%%%%%%%%%%%%%%%%%%%%%%%%%%%%%%%%%%%%%%%%%%%%%%%%%%%%%%%%%%%%
%%%% end fig 2
%%%%%%%%%%%%%%%%%%%%%%%%%%%%%%%%%%%%%%%%%%%%%%%%%%%%%%%%%%%%%%%%%%%%%%%
\begin{figure}[ptbh]
  \begin{center}
    \mbox{
      \epsfig{file=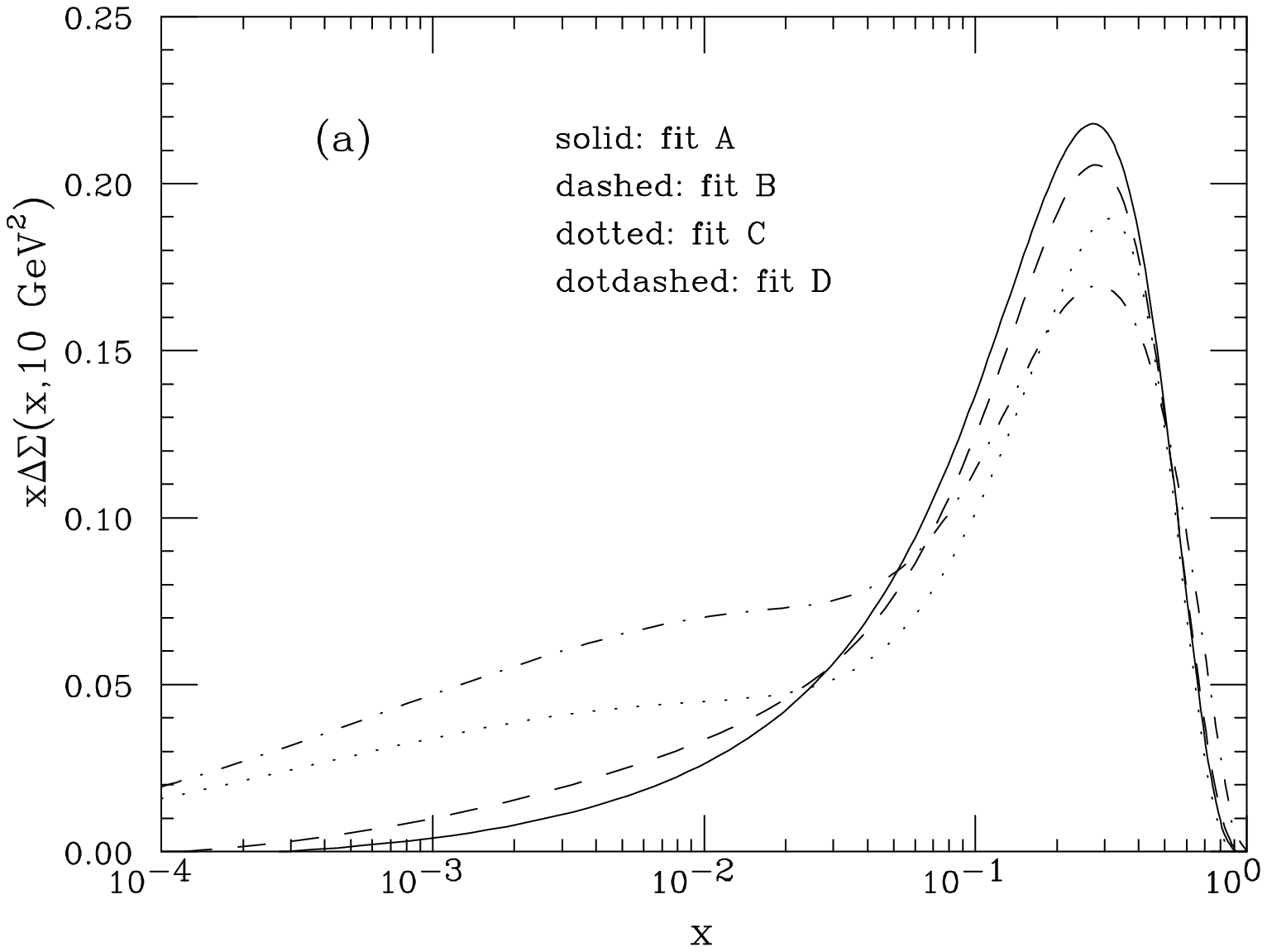,width=0.70\textwidth}
      }
\end{center}
\end{figure}
\begin{figure}[ptbh]
  \begin{center}
    \mbox{
      \epsfig{file=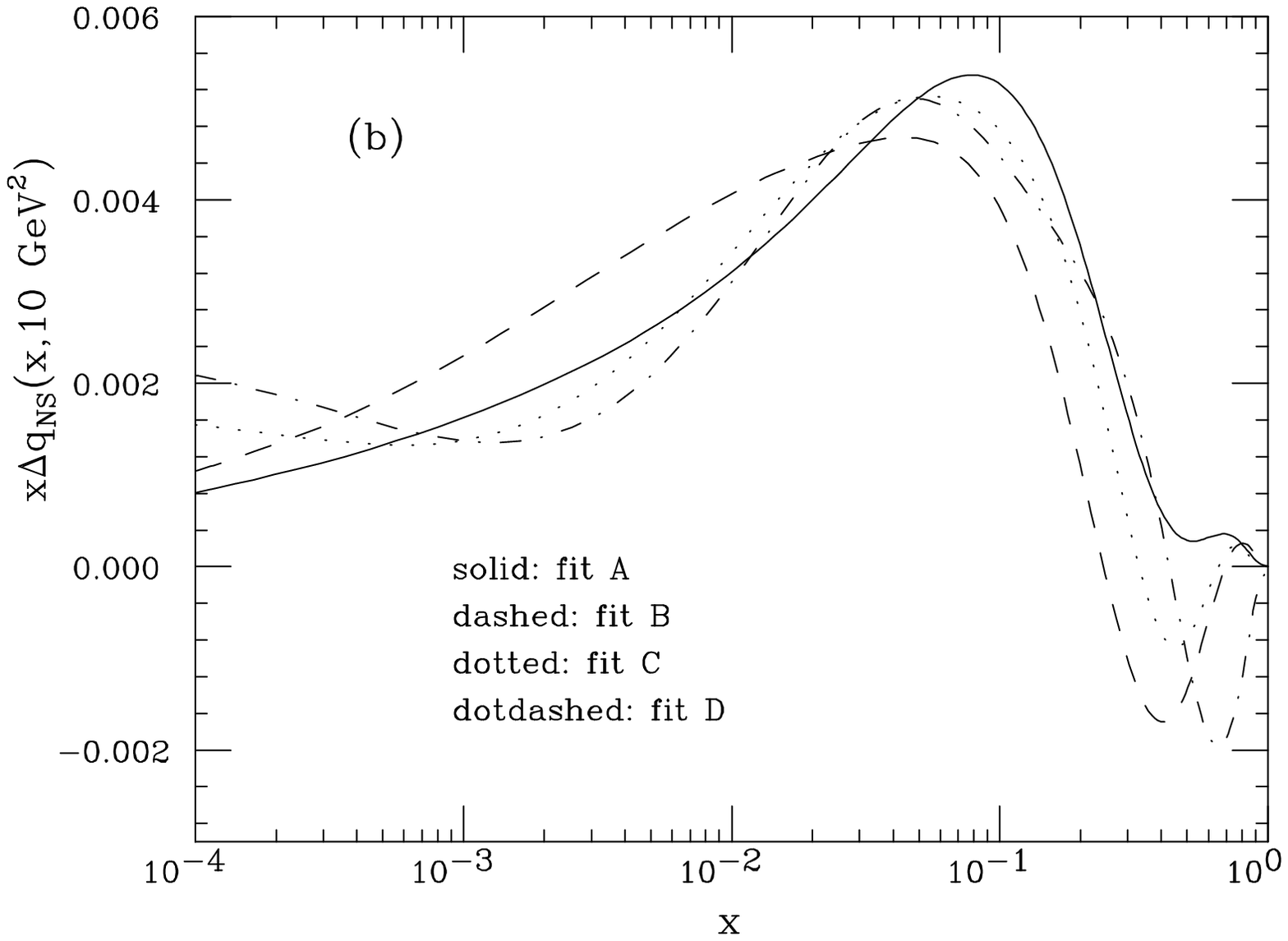,width=0.70\textwidth}
      }
  \end{center}
\end{figure}
\begin{figure}[ptbh]
  \begin{center}
    \mbox{
      \epsfig{file=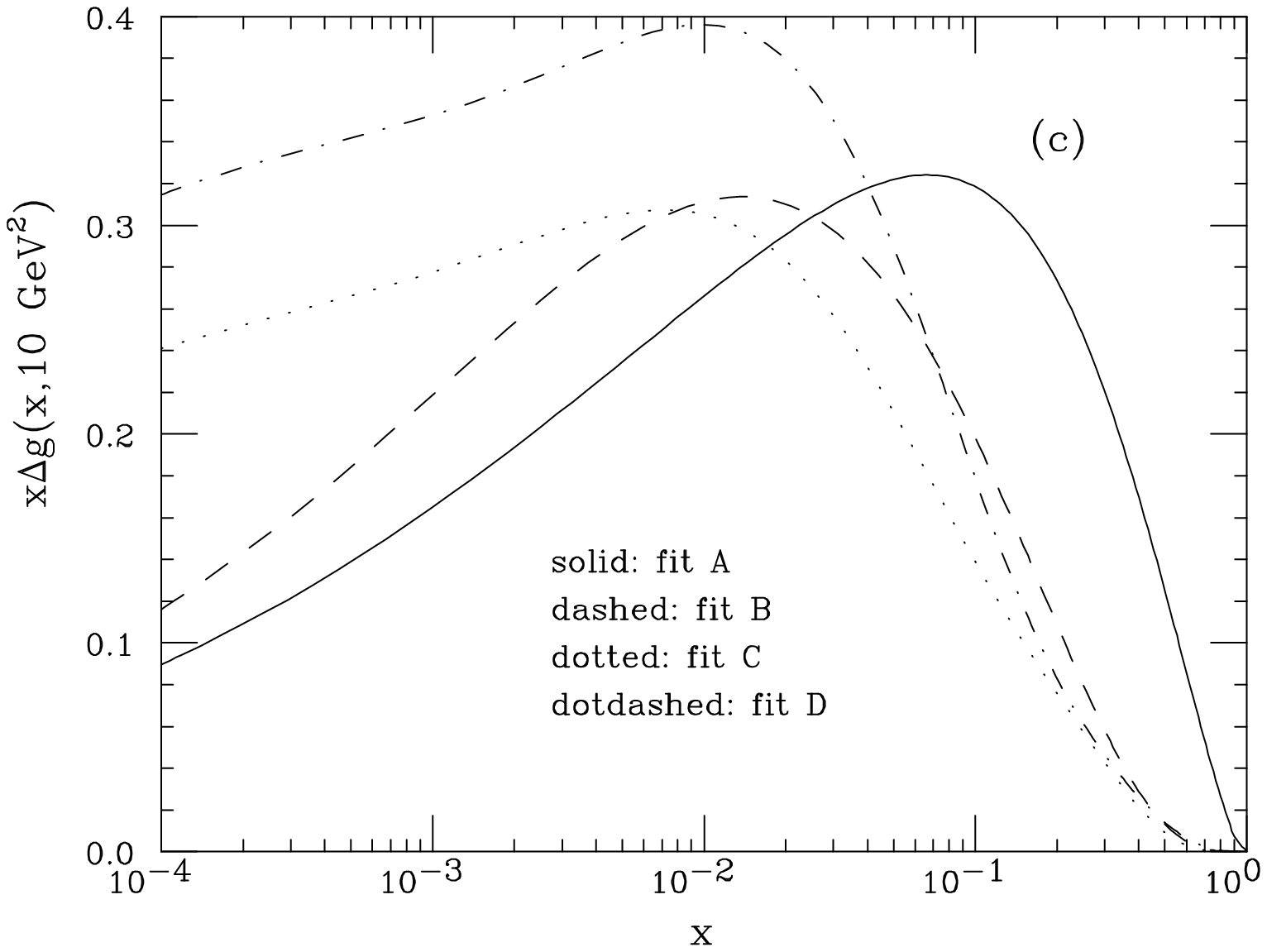,width=0.70\textwidth}
      }
  \ccaption{}{\label{dg}
Polarized quark singlet (a) nonsinglet (b) and gluon (c) distributions
for fits A--D at $Q^2=10$~GeV$^2$.
}
  \end{center}
\end{figure}
%%%%%%%%%%%%%%%%%%%%%%%%%%%%%%%%%%%%%%%%%%%%%%%%%%%%%%%%%%%%%%%%%%%%%%%


\begin{thebibliography}{99}
\baselineskip14pt
\bibitem{revs} G.~Altarelli, in ``The challenging questions'', Proc.
of the 1989 Erice School, A.~Zichichi, ed. (Plenum, New York, 1990);\\
R.D.~Ball, in ``The Spin Structure of the Nucleon'', Proc. of the 1995
Erice School of Nucleon Structure, B. Frois and V.W. Hughes, ed.
{\tt hep-ph/9511330};\\
S.~Forte, in the proceedings of the ``14th International Conference on
Particles and Nuclei (PANIC96)'' and ``12th International Symposium on
High Energy Spin Physics (SPIN96)'', {\tt hep-ph/9610238};\\
G.~Ridolfi, in the proceedings of the ``International Workshop on Deep
Inelastic Scattering and Related Phenomena (DIS96)'', {\tt hep-ph/9610214}.
\bibitem{EMC} EMC Collaboration, J.~Ashman \etal, \NP\vyp{B328}{1989}{1}.
\bibitem{SMCp}SMC Collaboration, D.~Adams \etal, \PL\vyp{B329}{1994}{399}.
\bibitem{E143p} E143 Collaboration, K.~Abe \etal, \PRL\vyp{74}{1995}{346}.
\bibitem{E143new} E143 Collaboration,  K.~Abe \etal, \PL\vyp{B364}{1995}{61}.
\bibitem{SMCd}SMC Collaboration, D.~Adams \etal, \PL\vyp{B357}{1995}{248}.
\bibitem{E143d} E143 Collaboration, K.~Abe \etal, \PRL\vyp{75}{1995}{25}.
\bibitem{SMCdnew} SMC Collaboration,  D.~Adams \etal, {\tt  hep-ex/9702005}.
\bibitem{E142} E142 Collaboration, P.L.~Anthony \etal,
\PR\vyp{D54}{1996}{6620}.
\bibitem{E154} Y.G.~Kolomensky for the E154 Collaboration, talk at the
``12th International Symposium on High Energy Spin Physics (SPIN96)''.
\bibitem{NLO} R.~Mertig and W.~L.~van~Neerven, \ZP\vyp{C70}{1996}{637};\\
W.~Vogelsang, \PR\vyp{D54}{1996}{2023}.
\bibitem{revHera} R.D.~Ball and A.~DeRoeck, in the proceedings of
the ``International Workshop on Deep Inelastic Scattering and Related
Phenomena (DIS96)'', {\tt hep-ph/9609309}, and ref. therein.
\bibitem{Bj} J.D.~Bjorken, \PR\vyp{148}{1966}{1467}.
\bibitem{EK93} J.~Ellis and M.~Karliner, \PL\vyp{B313}{1993}{131}.
\bibitem{AP} G.~Altarelli and G.~Parisi, \NP\vyp{B126}{1977}{298}.
\bibitem{BFRa} R.D.~Ball, S.~Forte and G.~Ridolfi, \NP\vyp{B444}{1995}{287}.
\bibitem{BFRb} R.D.~Ball, S.~Forte and G.~Ridolfi, \PL\vyp{B378}{1996}{255}.
\bibitem{closeroberts} F.~E.~Close and R.~G.~Roberts, \PL\vyp{B336}{1994}{257}.
\bibitem{Heimann} R.~L.~Heimann, \NP\vyp{B64}{1973}{429}.
\bibitem{DeRuj} A.~De~R\'ujula et al., \PR\vyp{10}{1974}{1649}.
\bibitem{BF} R.D.~Ball and S.~Forte, \PL\vyp{B335}{1994}{77};
\vyp{B336}{1994}{77}; \APP\vyp{B26}{1995}{2097}.
\bibitem{prise} M.~A.~Ahmed and G.~G.~Ross, \PL\vyp{B56}{1975}{385};\\
M.~B.~Einhorn and J.~Soffer, \NP\vyp{B74}{1986}{714};\\
A.~Berera, \PL\vyp{B293}{1992}{445}.
\bibitem{EK} J.~Ellis and M.~Karliner, \PL\vyp{B341}{1995}{397};
talk in ``The Spin Structure of the Nucleon'', Proc. of the 1995
Erice School of Nucleon Structure, B. Frois and V.W. Hughes, ed.
({\tt hep-ph/9601280}).
\bibitem{polpart} T.~Gehrmann and W.~J.~Stirling, \PR\vyp{D53}{1996}{6100};\\
M.~Gl\"uck et al., \PR\vyp{D53}{1996}{4775}.
\bibitem{altarelliross} G.~Altarelli and G.~G.~Ross, \PL\vyp{B212}{1988}{391}.
\bibitem{efremov} A.V.~Efremov and O.V.~Teryaev, Dubna preprint
E2-88-287 (unpublished).
\bibitem{carlitz}R.D.~Carlitz, J.C.~Collins
and A.H.~Mueller, \PL\vyp{B214}{1988}{229}.
\bibitem{AL} G.~Altarelli and B.~Lampe, \ZP\vyp{C47}{1990}{315}.
\bibitem{FBR} S.~Forte, R.D.~Ball and G.~Ridolfi, in the proceedings of
the ``International Workshop on Deep Inelastic Scattering and Related
Phenomena (DIS96)'', {\tt hep-ph/9608399}.
\bibitem{BFKL} L.N.~Lipatov, {\it Sov. J. Nucl. Phys.} {\bf 23} (1977) 338;\\
E.A.~Kuraev, L.N.~Lipatov and V.S.~Fadin,
{\it Sov. Phys. JETP} {\bf 45} (1977) 199;\\
Ya.~Balitskii and L.N.~Lipatov, {\it Sov. J. Nucl. Phys.} {\bf 28} (1978) 822.
\bibitem{KL} R.~Kirschner and L.~Lipatov, \NP\vyp{B213}{1983}{122}.
\bibitem{BER} B.I.~Ermolaev, S.I.~Manaenkov and M.G.~Ryskin,
\ZP\vyp{C69}{1996}{259};\\
J.~Bartels, B.~I.~Ermolaev and M.~G.~Ryskin,
\ZP\vyp{C70}{1996}{273}; {\tt hep-ph/9603204}.
\bibitem{NMCF2} NMC Collaboration, M.~Arneodo
\PL\vyp{B364}{1995}{107}.
\bibitem{SLACR} L.~W.~Whitlow et al.,
\PL\vyp{B250}{1990}{193}.
\bibitem{corrnuc} L.~L.~Frankfurt and M.~Strikman,
{\it Nucl. Phys.} {\bf A405} (1983) 557;\\
J.~L.~Friar et al., \PR\vyp{C42}{1990}{2310};\\
C.~Ciofi degli Atti et al., \PR\vyp{C48}{1993}{968}.
\bibitem{alf} S.~Bethke, talk at the ``High-energy Physics International
Euroconference on Quantum Chromodynamics (QCD 96)'', {\tt hep-ex/9609014};\\
G.~Altarelli, talks at the ``NATO Advanced Study Institute
on Techniques and Concepts of High-Energy Physics'' and at the ``Cracow
International Symposium on Radiative Corrections (CRAD 96)'',
{\tt hep-ph/9611239};\\
P.N.~Burrows, talk at the ``Cracow International Symposium on Radiative
Corrections (CRAD 96)'', {\tt hep-ph/9612007}.
\bibitem{CRb} F.~E.~Close and R.~G.~Roberts, \PL\vyp{B316}{1993}{165}.
\bibitem{su3}  B.~Ehrnsperger and A.~Sch\"afer,
\PL\vyp{B348}{1995}{619};\\
J.~Lichtenstadt and H.~J.~Lipkin,
\PL\vyp{B353}{1995}{119};\\
J.~Dai et al., {\it Phys. Rev.} {\bf D53} (1996) 273;\\
P.~G.~Ratcliffe, {\it Phys. Lett.} {\bf B365} (1996) 383.
\bibitem{ht} See L.~Mankiewicz, E.~Stein and A.~Sch\"afer,
{\tt hep-ph/9510418} and ref. therein.
\bibitem{PDG} Particle Data Group, \PR\vyp{D54}{1996}{1}.
\bibitem{AlRi} G.~Altarelli and G.~Ridolfi, \NPBPS\vyp{39B}{1995}{106}.
\bibitem{EJ} J.~Ellis and R.~L.~Jaffe, \PR\vyp{\bf D9}{1974}{1444}.
\bibitem{instantons} S.~Forte, \PL\vyp{B224}{1989}{189};
\NP\vyp{B331}{1990}{1};\\
G.M.~Shore and G.~Veneziano, \PL\vyp{B244}{1990}{75};
\NP\vyp{B381}{1992}{23};\\
R.D.~Ball, \PL\vyp{B266}{1991}{473}.
\bibitem{NNNLO} S.~G.~Gorishny and S.~A.~Larin, \PL\vyp{B172}{1986}{109};\\
S.~A.~Larin and J.~A.~M.~Vermaseren, \PL\vyp{B259}{1991}{345}.
\bibitem{HERAS} R.D. Ball et al, in the proceedings of the
`Workshop on Future Physics at HERA', Hamburg, {\tt hep-ph/9609515}.
\vskip12pt
\end{thebibliography}
\end{document}